\documentclass[showpacs,preprintnumbers,amsmath,amssymb,superscriptaddress,nofootinbib,english,floatfix]{revtex4}

\usepackage{graphicx}
\usepackage{dcolumn}
\usepackage{bm}
\usepackage{epsfig}
\usepackage{graphicx}
\usepackage{hyperref}
\usepackage[usenames]{color}
\usepackage{url}

\hypersetup{
    colorlinks=true,
    linkcolor=green,
    citecolor=blue,
}

\newcommand{\remove}[1]{}

\def\be{\begin{equation}}
\def\ee{\end{equation}}
\def\ba{\begin{eqnarray}}
\def\ea{\end{eqnarray}}

\begin{document}

\title{Detection Prospects for Solar and Terrestrial Chameleons}

\author{Philippe~Brax}
\email[Email address: ]{philippe.brax@cea.fr}
\affiliation{Institut de Physique Theorique, CEA, IPhT, CNRS, URA 2306, F-91191Gif/Yvette Cedex, France}

\author{Axel  Lindner}
\affiliation{Deutsches Electronen-Synchrotron, Notkestr. 85, 22607 Hamburg.}
\email{axel.lindner@desy.de}

\author{Konstantin~Zioutas}
\affiliation{University of Patras, GR 26504 PATRAS, Greece}
\email{zioutas@cern.ch}

\date{\today}

\begin{abstract}
Dark energy models, such as the chameleon,  where the acceleration of the expansion of the universe results from the dynamics of a scalar field coupled to matter, suffer from the potential existence of a fifth force. Three known mechanisms have been proposed to restore General Relativity in the solar system and the laboratory, which are the symmetron/Damour-Polyakov effect, the Vainshtein  property and  the chameleon screening.  Here, we propose to probe the existence of chameleons in the laboratory,   considering  their particle physics consequences. We
envisage the resonant and non-resonant production of chameleons in the sun and their back-conversion into X-ray photons in a solar helioscope pipe  such as the one used by CAST. We find that the resonant contribution is only present for a very small range of chameleon couplings to matter and occurs in  extremely narrow magnetised regions in the sun. However, in the non-resonant case, chameleons are created in all available solar magnetic regions and for a very wide range of chameleon couplings to photons and matter.
Such chameleons could leave the sun and be back-converted  into X-ray photons above the photosphere or in the magnetic pipe of a helioscope. A detection of these X-rays would indicate the existence of chameleons.
The number of produced chameleons and regenerated X-ray photons varies depending on the strength of the solar magnetic field in the production shell,  the matter  and photon  couplings of chameleons, and the magnetic field length and magnitude inside a dipole magnet. We focus on a template model for the solar magnetic field: a constant magnetic field  in a narrow shell surrounding the tachocline. The X-ray photons in a helioscope pipe obtained from back-conversion of the  chameleons created inside  the sun have a  spectrum which is peaked in the sub-keV region, just below the actual  sensitivity range of the present axion helioscopes. Nevertheless they are  detectable by present day magnetic helioscopes like CAST and Sumico, which were built originally for solar axions.   We also propose a chameleon-through-a-wall experiment whereby  X-ray photons from a synchroton radiation source could be converted into chameleons inside a dipole magnet, then pass a  wall which is opaque to X-rays before being back-converted into X-ray photons in a second magnet downstream. We show that this could provide a direct signature  for the existence of chameleon particles.

\end{abstract}

\maketitle

\section{Introduction}
The acceleration of the expansion rate of the Universe is still a deep mystery despite more than ten years of intense activity both experimentally\cite{Riess:1998cb,Perlmutter:1998np} and theoretically\cite{Copeland:2006wr,Brax:2009ae}. If the acceleration of the Universe is due to an unknown form of matter, dark energy, it comprises more than seventy percent of the energy budget of the
Universe. Together with dark matter, some ninety five percent of the energy contents of the Universe are unknown to us and only detected through indirect cosmological and gravitational astrophysical probes.

Of course, the situation would be greatly clarified if some imprint of the existence of dark energy could be discovered experimentally. In this sense, looking for dark energy can be put in parallel with the search for dark matter (see \cite{silk} and references therein).
For dark matter, the detection methods  comprise the identification of potential candidates in the laboratory (ranging from experiments at the LHC looking for very heavy new particles to light-shining-through-a-wall experiments at the opposite end of the mass scale) as well as finding signatures in astrophysical environments like the production of axions in the sun for example.
Even if dark matter particles were found, dedicated efforts would be required to ascertain that these candidates do act as  the pervading astrophysical dark matter.
Hence  experimental searches for dark matter in indirect (by looking for annihilation products of dark matter particles for example) or direct channels are also  under scrutiny.
The latter includes scattering experiments mainly focusing on heavy WIMP dark matter in the multi-10 GeV range and axion dark matter searches in the $\mu$eV mass regime.

Searches for dark energy clearly lag behind their dark matter counterparts. A particular reason for this is the lack of particle candidates as well as realistic proposals for smoking-gun detections of dark energy.
Here we propose astrophysical and laboratory experiments to probe a particular class of dark energy models, i.e. the chameleons\cite{Khoury:2003aq,Khoury:2003rn,Brax:2004qh,Mota:2006fz}, and their distinctive environmentally dependent features.
It should be stressed that even with a clear identification of chameleons in the laboratory, this would not provide a direct explanation to  dark energy itself. Specialised experiments would have to be pursued in order to prove that chameleons correspond to  the dark energy of  the Universe.
However, it would clearly be a big step forward in our understanding of the Universe if candidates for dark matter and dark energy were identified in the laboratory.

If dark energy is a new form of matter, strong candidates could be scalar fields, which have not been detected so far, interacting with both matter and photons. Indeed, scalar fields can provide the potential energy which would drive the acceleration of the expansion rate of the Universe. Unfortunately this implies that the scalar field mass on cosmological scales is of the order of the Hubble rate now, i.e. $H_0\sim 10^{-33}$ eV. Such a tiny mass would lead to large gravitational effects and the presence of a long range fifth force. The Cassini bound on the coupling of such scalar fields to matter\cite{Bertotti:2003rm}, together with other gravitational tests in the laboratory\cite{Will:2001mx},  lead to unnatural  models if a screening mechanism were not at play both in the solar system and the laboratory. Fortunately, known screening mechanisms exist  such as the thin shell effect for  chameleons and the Vainshtein mechanism for galileons, involving highly non-linear Lagrangians which can only be understood as low energy effective field theories\cite{Khoury:2003aq,Khoury:2003rn,Brax:2004qh,Mota:2006ed,Mota:2006fz,Hinterbichler:2010es,Brax:2010gi,Nicolis:2008,Olive:2007aj}. The origin of these models at higher energy is unknown. In this paper, we will focus on chameleons.

Chameleon models can describe  the acceleration of the expansion of the Universe on large scales while having a density dependent effective  mass which explains why they are screened in local experiments. Indeed, the rest mass of chameleons is large inside dense bodies implying that the field inside  does not extend further than a scale of order $m^{-1}$ where $m$ is the mass at a given point in the astrophysical body. Hence, the contributions to the value of the exterior field from inside the body are damped by a large Yukawa suppression $\exp(-m d)$, where $d$ is the distance from a point in the body to another one outside. Only a tiny shell at the surface of the body, called a thin shell, contributes to the field outside the body as the field is not Yukawa damped across this thin layer. Therefore the exterior of a body feels the field created only by a small shell and is thus small, i.e.  small enough to create a tiny modification of gravity and pass all the gravitational  tests on non-cosmological scales.

However, chameleons themselves can have detectable effects in the laboratory. Indeed, they can induce a small but detectable deviation from the Casimir force between two metallic plates \cite{Brax:2010xx}. They can also couple to two photons in  a Primakoff fashion\cite{primakoff} and may lead to observable phenomena  in optical experiments,  if, for example,   a laser beam goes across a magnetic region providing a bath of virtual photons\cite{pvlas}.  Using the same coupling to two photons, they could be produced in solar magnetic regions in a way akin to the axionic one\cite{sikivie}. The main difference between axions and chameleons would be the fact that chameleons couple to the photon polarisation perpendicular to a magnetic field (pseudo-scalars couple to the parallel polarisation), and the crucial variation of their effective  mass with the pressure-less energy density of the environment.  Given the pressure-less energy density profile,  the mass essentially depends on two parameters: an index $n$ which characterises the dark energy potential and the coupling constant  to matter $\beta$. Together with the coupling constant  to photons $\beta_\gamma$, this forms the three parameter family of chameleon models. Notice that for chameleons, the mass has no intrinsic meaning like for axions. Here the index $n$ and the two coupling constants  $\beta$ and $\beta_\gamma$ suffice.

If, in a dense magnetised plasma,  the chameleon mass matches the plasma frequency of the environment,  chameleons can be resonantly produced in the presence of a magnetic field. This could happen  in the sun in a narrow shell whose location depends on the effective rest mass of the chameleons.
On the other hand, non-resonant production is more widespread and occurs in all magnetised regions inside the sun.
Once created either resonantly or non-resonantly across large magnetic regions, chameleons essentially escape from the sun and reach earth where they can enter an helioscope and be back-converted into X-ray photons\cite{Brax:2010xq}. The fact that the emitted chameleons have a spectrum in the immediate sub-keV range helps them to overcome any barrier. Note that this only holds true as long as their mass in matter is smaller than the chameleon energy.
Low energy chameleons are reflected by dense matter. The chameleon emitted by the sun eventually reach earth where they can enter magnetic helioscopes like CAST\cite{CAST} and Sumico\cite{sumico} and be back-converted into X-ray photons. Additionally,  a small fraction of the solar chameleons could also be converted into X-ray photons in the photosphere and above, hence providing a possible explanation to the corona puzzle: the corona is hotter above magnetised solar surfaces. This reasoning has been used also for axions or axion-like particles in \cite{zzz} and applies to chameleons as well. In fact, as we show in this work, one can distinguish the photon emission coming from chameleons from that coming from axion conversion.

In order to estimate the sensitivity of possible experiments, we have applied present constraints mimicking the ones from  the CAST experiment and typical of helioscopes, i.e. a background noise of 1 event per hour in the 1-7 keV region, to draw the sensitivity  region in the chameleon parameter space at the 2$\sigma$ level. For the chameleon production inside the sun,  we have used the conservative assumption that the solar magnetic field is mainly effective in a shell of 0.01 solar radius around  the tachocline where a constant magnetic field of 30T is commonly  assumed to exist with this rather realistic width. With these hypotheses, we deduce both the chameleon spectrum and the regenerated photon spectra in a helioscope pipe, the photosphere and the corona. As we find that the photon spectrum is peaked below the main sensitivity region of the present helioscope detectors, these limits will be improved with future upgrades.

Constraints on the chameleon parameter space have already been given by the CHASE experiment at Fermilab using optical photons\cite{Upadhye:2009iv,Steffen:2010ze}. In particular for low values of the matter coupling, the coupling to photons is bounded from above $\beta_\gamma \le 10^{11}$. With the assumption concerning the production of chameleons in the sun, due to a shell of 0.01 solar radius ($R_\odot$) and a magnetic field of 30T, and a maximal chameleon emission of 10\% of the solar luminosity\cite{Gondolo:2008dd},  we obtain an energy loss bound on the photon coupling $\beta_\gamma \lesssim 10^{10.32}$. This is comparable to the CHASE bound obtained with optical photons.

However,  the results in optical cavities involve many reflections of the chameleons on the beamtube walls and thus are limited by the presence of possible  multiple scatterings. To avoid these uncertainties, we propose a new X-ray light-shining-through-a-wall experiment where these issues are absent. Similarly,
the possible detection of X-ray photons springing from the existence of solar chameleons would be a rather indirect way of studying the properties of chameleons. To overcome these obstacles, we suggest a new experiment where a photon flux would be injected in a magnetic pipe\cite{desy}. As a result, chameleons would be produced. If a wall is introduced after the first magnetic pipe, energetic chameleons would go through the wall and would regenerate photons on the other side downstream. Using the energy loss bound on the coupling of chameleons to photons, we find that a 5$\sigma$ detection  of photons would be within reach with an improved source of photons compared to the ones available now: a smoking gun for chameleons in the laboratory.

In the  first part of this paper, we recall general results on chameleons and their production in the sun. We then calculate the number of back-converted photons in a helioscope pipe. Finally we present a chameleon through wall experiment and  we conclude.

\section{Chameleons in the Sun}
\subsection{Chameleon models}

Chameleons are candidates to generate the recent acceleration of the expansion of the Universe. On large scales, the chameleon field takes a vacuum expectation value whose energy density drives the acceleration. On smaller scales, this field would be responsible for a fifth force. However, a screening mechanism in massive bodies allows the model to evade all the local gravitational tests.

Chameleon models depend on the shape of the dark energy potential. In the following we will always assume that $V(\phi)$ is a decreasing function. The coupling to matter $\beta$ is also crucial. Chameleons have an effective potential which becomes matter density dependent
\begin{equation}
V_{\rm eff}(\phi)= V(\phi) +e^{\beta \phi/m_{\rm  Pl}} \rho
\label{veff}
\end{equation}
where $\phi$ is the chameleon field and $\rho$ the non-relativistic matter density.
As the potential $V(\phi)$ decreases, the effective potential has always a matter dependent minimum.
Fluctuations around the vacuum expectation values at the minimum can be seen as
chameleon  particles which couple to matter.
The best constraint on $\beta$ comes from the energy levels of neutrons in the terrestrial gravitational field  and reads $\beta \lesssim 10^{11}$\cite{Brax:2011hb}.
The  density-dependent minimum is such that  the mass
of the scalar field becomes also density dependent. We will focus on inverse power law models defined by
\begin{equation}
V(\phi)= \Lambda^4+ \frac{\Lambda^{4+n}}{\phi^n}+ \dots
\end{equation}
where we have neglected higher inverse powers of the chameleon field. These models describe  chameleon properties with no loss of generality.
Moreover, although precise numerical results depend on $n$, generic properties are mostly independent of $n$. In the following we will focus on small values of $n$, i.e.
$n=1$ and $n=4$ when we investigate chameleon resonances in the sun and later only $n=4$ for the back-conversion of chameleons in the photosphere, the corona and helioscopes. A more thorough scanning of the values of $n$ should be envisaged when comparing our results to experimental data in helioscopes.

We will choose $\Lambda=2.4~ 10^{-12} {\rm GeV}$ to accommodate  the acceleration of expansion rate of the universe on large scales.
The potential has a minimum located at
\begin{equation}
\phi_{\rm min}=(\frac{n m_{\rm Pl} \Lambda ^{4+n}}{\beta
\rho})^{1/(n+1)}
\end{equation}
where $\rho$ is the total non-relativistic matter density. The chameleon rest  mass at the minimum is
\begin{equation}
m^2\approx \beta \frac{\rho}{m_{\rm Pl}} \frac{n+1}{\phi_{\rm min}} .
\label{mass}
\end{equation}
We will see that in a cylindrical pipe, this rest mass is only valid at large enough coupling to  matter $\beta$.
Chameleons also couple to photons in a way akin to the axion coupling
\begin{equation}
S_{EM}= -\int d^4 x \sqrt{-g} \frac{e^{\phi/M_\gamma}}{4} F^2
\end{equation}
where $g$ is the determinant of the metric $g_{\mu\nu}$ and $F^2= F_{\mu\nu}F^{\mu\nu}$ where $F_{\mu\nu}$ is the photon field strength.
This implies that the effective matter density in the effective potential is
\begin{equation}
\rho= \rho_m + \frac{\beta_\gamma}{\beta } \frac{B^2}{2}
\label{dense}
\end{equation}
where $B$ is the magnetic field. For instance, with $\beta=10^6$ and $\beta_\gamma=10^{10.29}$, the contribution of a magnetic field $B=30$ T, like the one around the tachocline, to the energy density is approximately equivalent to $7\cdot 10^{-8}\ {\rm g\cdot cm^{-3}}$, i.e. a negligible contribution to $\rho$ when $\rho_m={\cal O}(1) \ {\rm g\cdot cm^{-3}}$.
The chameleon parameter space depends on the discrete index $n$ and two continuous parameters $\beta=\frac{m_{\rm Pl}}{M_m}$, the coupling to matter, and $M_\gamma$ the suppression scale of the coupling to photons. It is also convenient to introduce the photon coupling parameter $\beta_\gamma =\frac{m_{\rm Pl}}{M_\gamma}$.

We have already mentioned different experimental and observational bounds on the coupling parameters of chameleons. From a theoretical point of view, very few constraints apply to the parameter space of chameleons. The shape of the potential which we have parameterised using the index $n$ is not restricted. We will only invoke a naturality criterion and assume that $n$ is not a large number although we are not aware  of any deep physical reason for this. When it comes to the coupling constants to matter and photons, we will simply impose that the couplings are greater than one, e.g. corresponding to a scale $M_\gamma \le m_{\rm Pl}$. Again, this type of restriction is only due to the fact that chameleon theories must be viewed as  effective field theories only valid below a scale which could be smallest value between $M_m$ and $M_\gamma$. Of course, as the high energy description of chameleon theories valid up to very large scales, maybe the Planck scale, is not known, we cannot be too strict with this type of argument.

On the other hand, when the chameleons are conformally coupled to matter fields, e.g. fermions, and that one studies the low energy physics obtained after integrating out the heavy fields, the coupling constants $\beta$ and $\beta_\gamma$ can be related\cite{burrage}
\begin{equation}
\beta_\gamma =(3 N_f + \frac{N_f^+}{3})\frac{\alpha\beta }{4}
\end{equation}
where $\alpha$ is the fine structure constant.
In this case, their relationship depends on the total number of fermions $N_f$, due to a conformal anomaly, and the number of heavy fermions $N_f^+$.
This would naturally lead to a hierarchy between $\beta_\gamma \ll \beta$ when a large number of fermions is present.
In practice and in the absence of knowledge about the number of fermions in the theory at high energy and whether chameleons are conformally coupled in this high energy regime, we have chosen to use $\beta$ and $\beta_\gamma$ as independent parameters.

\subsection{Chameleon-photon oscillations}

Chameleons can be produced during the propagation of photons inside  the macroscopic magnetic fields of the  sun. These chameleons would then escape from the sun and reach the earth, where they could be detected
by helioscopes  such as CAST.
Indeed chameleons interact with matter with a coupling constant depending on the fermion mass and the combination $g_f= \beta \frac{m_f}{m_{\rm Pl}}$. Interactions are therefore  stronger with protons than with electrons in the solar plasma.
The interaction rate with a fermions can be estimated as
\begin{equation}
\Gamma_f \sim n_f\frac{g_f^4}{m_f^2}
\end{equation}
Taking the proton rate as the dominant part and using electroneutrality, we find that the mean free path $\lambda_p=\frac{1}{\Gamma_p}$ reads
\begin{equation}
\lambda_p \sim \beta^{-4} \frac{m_{\rm Pl}^4}{m_p \rho}
\end{equation}
where $\rho$ is the solar density and $m_p$ the proton mass. The mean free path is shorter in dense regions and reads
\begin{equation}
\frac{\lambda_p}{R_\odot} \sim 10^{21}(\frac{10^{11}}{\beta})^4 (\frac{100 {\rm g\cdot cm^{-3}}}{\rho})
\end{equation}
which far exceeds the size of the sun. Hence solar  chameleons escape through the solar plasma unscathed.

The only possible interaction of chameleons is with photons in a magnetised region. The mixing of photons and chameleons depends on
\begin{equation}
k^2(\omega )=\omega^2- m^2_{\rm eff} (\frac{\cos \theta +1}{2\cos 2\theta})
\label{k}
\end{equation}
where the effective mass is (see (\ref{mass}))
\begin{equation}
m^2_{\rm eff}=m^2-\frac{B^2}{M_\gamma^2}-\omega^2_{\rm pl}
\label{eff}
\end{equation}
and
$\omega$ is the initial frequency of the incoming photons. This depends on
the mixing angle which is given by
\begin{equation}
\tan 2\theta= \frac{2\omega B}{M_\gamma m^2_{\rm eff}}
\label{theta}
\end{equation}
and the plasma frequency which is
\begin{equation}
\omega^2_{\rm pl}= \frac{4\pi \alpha n_e}{m_e}
\end{equation}
where $\alpha$ is the fine structure constant. For instance, when $\omega_{\rm pl}=0.8 \ {\rm eV}$ and $\beta_\gamma=10^{10.29}$, the magnetic contribution
to the mass for $B=10$ T is $10^{-14}$ times smaller than the plasma frequency.
Electro-neutrality implies that,  in the sun, $n_e= \frac{\rho_m}{m_p}$, where $m_p$ is the proton mass.
When $\theta\ll 1$: as is the case in most helioscope pipes and also inside the sun,  we have
\begin{equation}
\theta= \frac{\omega B}{M_\gamma m^2_{\rm eff}}.
\end{equation}
Chameleon production in a magnetised region of size $L$ is obtained from the transition probability
\begin{equation}
P_{\rm chameleon}(\omega)= \sin^2 (2\theta) \sin^2 (\frac{\Delta}{\cos 2\theta})
\label{prob}
\end{equation}
where $\Delta= m^2_{\rm eff} L/4\omega$. For instance with $BL\approx 90 \ T.m$ and $\beta_\gamma=10^{10.29}$, the dependence of this conversion probability
on the photon energy $\omega$ can be seen on figure 6 where typical values of order $10^{-13}$ are obtained.

\begin{figure}
\begin{center}
\includegraphics[width=7cm]{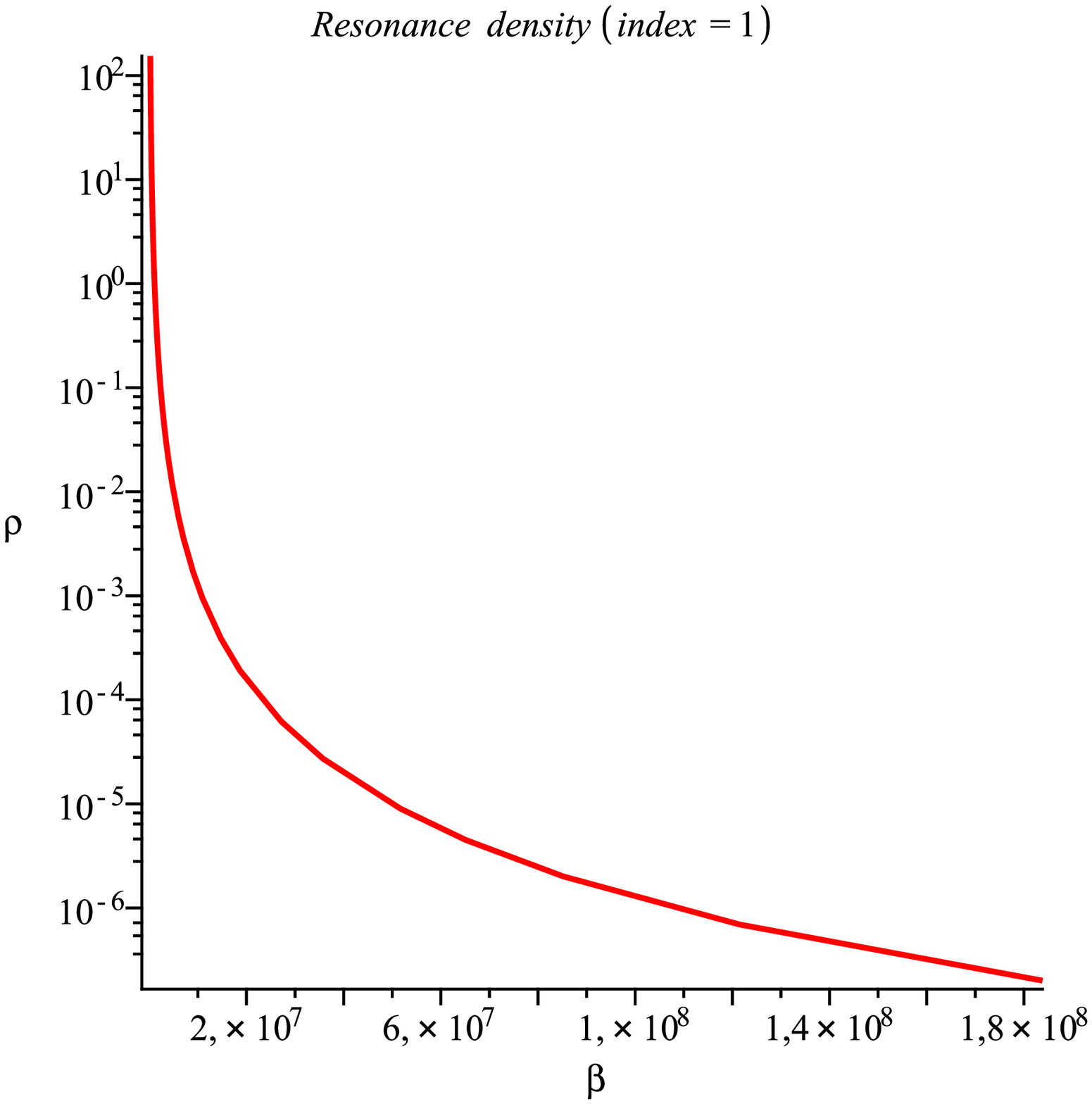}\hspace*{1cm}
\includegraphics[width=7cm]{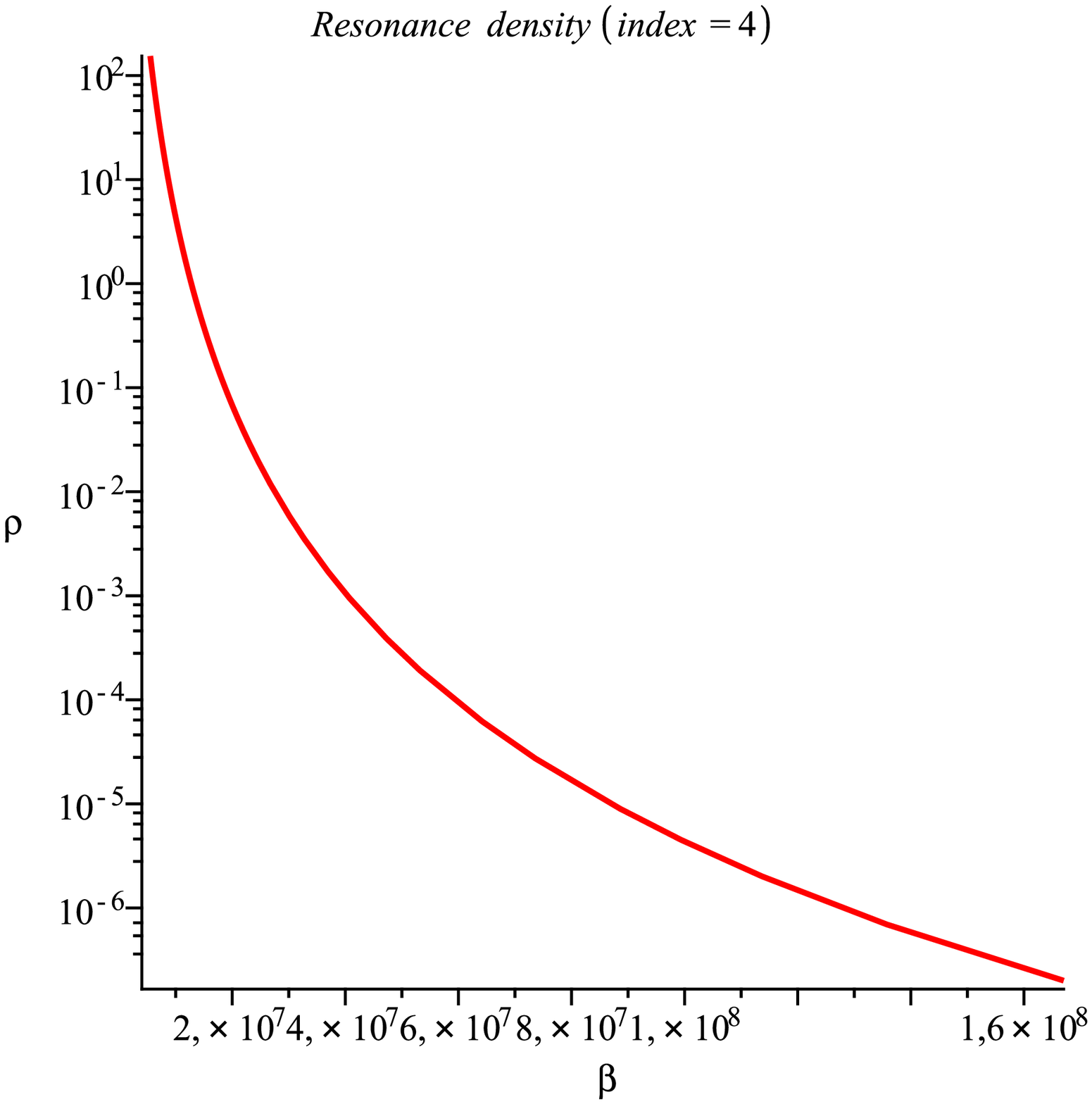}

\caption{The matter  density at the resonance in $\rm g\cdot cm^{-3}$ as a function of the matter coupling constant $\beta$ for $n=1$ and $n=4$. Notice that only a limited range of $\beta$ can lead to a resonance in the sun. Nuclear physics constraints always impose $\beta \lesssim 10^{11}$\cite{Brax:2011hb}.}
\end{center}
\end{figure}

\subsection{Chameleon propagation in a dense medium}

It is  important to notice that chameleon production is only possible when $\omega^2\ge m^2_{\rm eff}$ (see (\ref{k})) for small values of $\theta$. We will find that for large values of $\beta$, this condition is hard to meet in the sun. Neglecting the magnetic contribution to the mass\footnote{For $\beta=10^6$ and $\beta_\gamma=10^{10.29}$, the magnetic contribution to the energy density is $10^{-11}$ times smaller than the matter density close to the tachocline and only around  a few percent in the sparsest regions of  the sun's atmosphere.} , we have the following expression for the effective  mass (\ref{eff})
\begin{equation}
m^2_{\rm eff}= \beta^{1+1/(n+1)} \omega_\rho^2  - \omega^2_{\rm Pl}
\end{equation}
where we have introduced the density dependent frequency
\begin{equation}
\omega_\rho^2= \frac{(n+1) \rho}{m_{\rm Pl}}(\frac{\rho}{nm_{\rm Pl} \Lambda^{n+4}})^{1/(n+1)}
\end{equation}
For large enough $\beta$, the effective mass increases like $m^2_{\rm eff}\approx  \beta^{1+1/(n+1)} \omega_\rho^2$. This implies that low energy chameleons are attenuated in the sun. Chameleons only propagate when
\begin{equation}
\beta \le (\frac{\omega}{\omega_\rho})^{2(n+1)/(n+2)}
\end{equation}
Let us consider the case of chameleon production in a narrow magnetic region, say around the tachocline $R\sim 0.7 R_\odot$. As $\omega_\rho$ decreases with decreasing $\rho$, once a chameleon of energy $\omega$ has been created, it can propagate all the way to the outer sun. Denoting by $\rho_{\rm pro}$ the density in the production region and
$\omega_{\rm pro}= \omega_{\rho_{\rm pro}}$, a bound on the range of allowed couplings is
\begin{equation}
\beta \le (\frac{\omega}{\omega_{\rm pro}})^{2(n+1)/(n+2)}
\end{equation}
For a given $\beta$, we then see that energetic chameleons can always satisfy this bound while chameleons of lower energy are attenuated in the sun. On the other hand if we are interested in back-converted photons from chameleons in a magnetic helioscope with an energy range, e.g. $1\ {\rm keV}\le \omega  \le 7\  {\rm keV}$, we find that chameleons can only be produced if $\beta$ is smaller than an upper bound $\beta_{\rm max}$. The value of $\beta_{\rm max}$ depends on the density in the production region and the index $n$.
For instance, for $n=4$ and a production around the tachocline, we find that $\beta\lesssim 10^{11}$ for the propagation of chameleons in the detectable energy range $1 \ {\rm keV}\le \omega \le 7 \ {\rm keV}$. This is a very wide range of couplings which is compatible with the bound from the neutron energy levels in the terrestrial gravitational field.

\subsection{The chameleonic journey from the tachocline to a helioscope}

We will estimate the production of chameleons in magnetic regions in the sun. We will focus on a rather conservative assumption, where a constant magnetic field is present in a small shell around the tachocline about 0.7 $R_\odot$ (the width of the tachocline is around $0.04 \ R_\odot$ with a magnetic field in the 20-50T range). In this region, the production can be of two types. The first one is due to the presence of a resonance where the effective mass vanishes. This leads to the creation of chameleons over a narrow radial strip of extremely small width. The second type is the non-resonant production over the whole width of the magnetic region. Once created, chameleons interact very little with matter and can only occasionally be back-converted to photons. If this  happens deep inside the sun, this represents a tiny modification of the solar radiative transfer, but it  can be safely neglected. If this chameleon to photon back-conversion takes place in the photosphere and above where the photon mean free path is large, the resulting back converted photons would escape from the sun and have a spectrum which could contribute to  the sun's soft  X-ray luminosity. Interestingly,  the sun's X-ray luminosity far exceeds the expected one for a black body at T= 5780 K. This excess radiation reflects the puzzling corona problem,  first noticed in 1939\cite{corona}. Remarkably, the corona is indeed hotter above magnetised regions like sunspots. Chameleons leaving the sun arrive unhindered in the earth's atmosphere where they penetrate due to their high energy. They also go through matter and can be back converted into photons in magnetic helioscopes. We will describe the production mechanisms in the next section.

\section{Resonant and Non-resonant Solar Chameleon Production}

\subsection{Resonant production}
Let us consider the resonant case when the effective mass (\ref{eff}) vanishes somewhere inside the sun\cite{etzel}. Notice that the actual rest mass of the chameleon does not vanish at the resonance. Of course, such a resonant effect is only of interest for this work if it occurs in a magnetised region of the sun.
Neglecting as always the magnetic contribution to the mass we have
\begin{equation}
m^2_{\rm eff}=  \frac{4\pi \alpha \rho_{\rm res}}{m_e m_p}((\frac{\rho}{\rho_{\rm res}})^{1+\frac{1}{n+1}}- \frac{\rho}{\rho_{\rm res}})
\end{equation}
where we have introduced
\begin{equation}
\rho_{\rm res}= \frac{nm_{\rm Pl} \Lambda^{4+n}}{\beta} (\frac{4\pi \alpha m_{\rm Pl}}{\beta (n+1) m_e m_p})^{n+1}
\end{equation}
The resonance happens in the sun when $\rho(R_{\rm res})= \rho_{\rm res}$. This depends on $\beta$ and $n$.
At the resonance we have $\theta= \frac{\pi}{4}$ and $\sin^2 (2\theta)=1 $.

In the vicinity of the resonance, the mixing angle is given by
\begin{equation}
\cos 2\theta\approx \frac{M_\gamma}{ 2 \omega B_{\rm res}} m^2_{\rm eff}
\end{equation}
where $B_{\rm res}$ is the magnetic field at the resonance, say the assumed $B$ field in the tachocline. The width of the resonance $\delta l$  is obtained when $\cos 2 \theta \sim 1$. To linear order this implies that
\begin{equation}
\delta l= \frac{2 \omega B_{\rm res}}{M_\gamma} \frac{dr}{d\rho}\vert_{\rm res} \frac{(n+1) m_e m_p}{ 4\pi \alpha}
\label{wi}
\end{equation}
depending on the derivative $\frac{dr}{d\rho}$ at the resonance.
As long as  $B\delta l/M_{\gamma} \ll 1$  the production probability through one resonance is
\begin{equation}
P_{\rm cham}(\omega) \approx  \frac{1}{4} \frac{\delta l^2 B^2_{\rm res}}{M^2_\gamma}
\end{equation}
which can be conveniently written as
\begin{equation}
P_{\rm cham}(\omega) \approx  \frac{\omega^2}{\omega_{\rm cham}^2}
\end{equation}
where
\begin{equation}
\omega^2_{\rm cham}= \frac{M^4_\gamma}{B^4_{\rm res}} (\frac{d\rho}{dr}\vert_{\rm res})^2 (\frac{4\pi \alpha}{(1+n)m_e m_p})^2
\end{equation}
\begin{figure}
\begin{center}
\includegraphics[width=7cm]{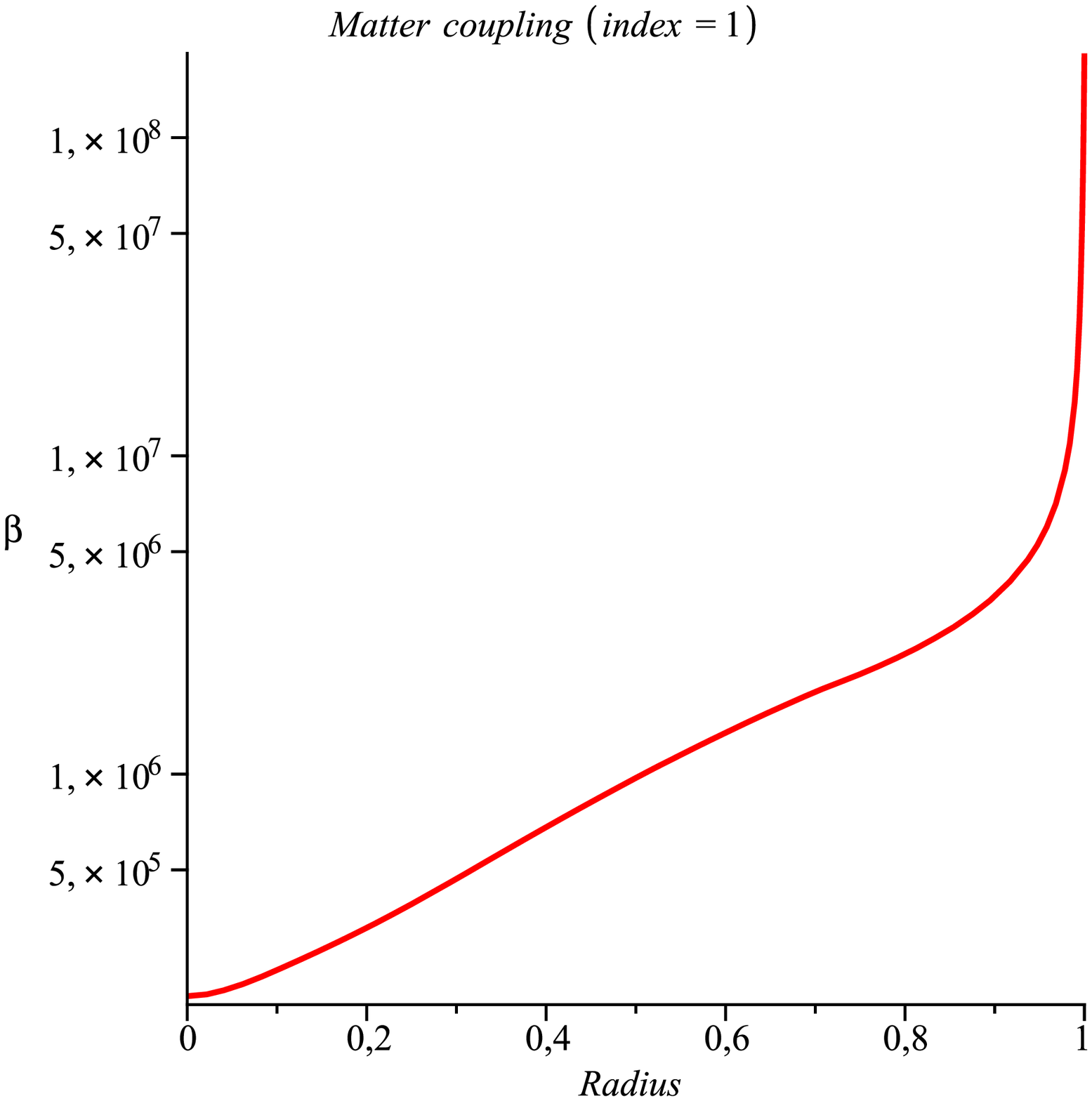}\hspace*{1cm}
\includegraphics[width=7cm]{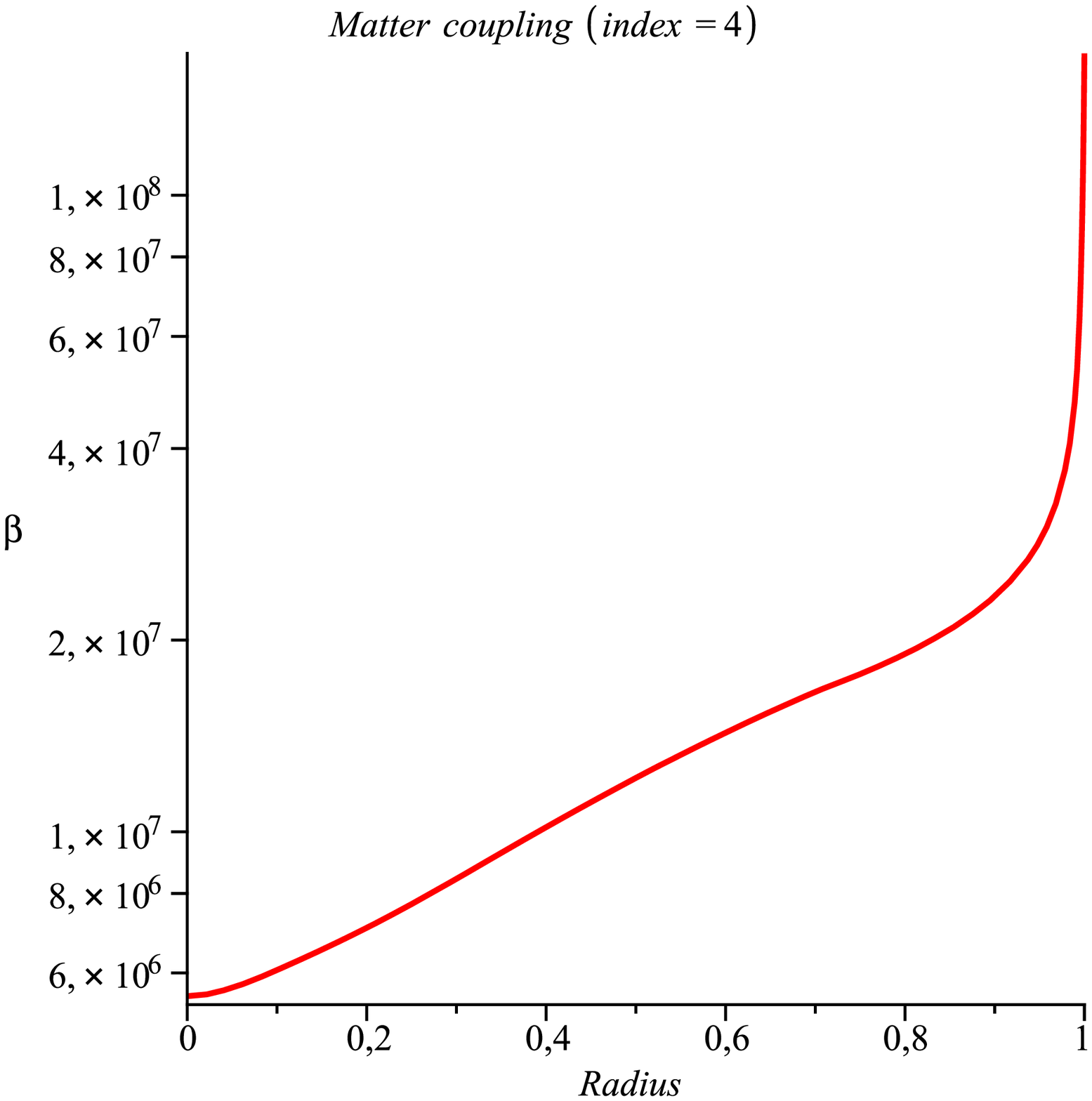}

\caption{The matter coupling constant $\beta$ leading to a resonance as a function of the radial distance of the resonance (in units of the solar radius) when $n=1$ and $n=4$.  Nuclear physics constraints impose $\beta \lesssim 10^{11}$\cite{Brax:2011hb}.}
\end{center}
\end{figure}
Thermal photons in the sun perform a random walk.
In one second,  photons scatter
${\cal N}= \frac{c}{\lambda}$ times on average. Moreover,
in one second, a photon covers a distance $d= \sqrt{\cal N} \lambda$ on average. As a result, the number of times that one photon goes through the resonance in one second is
$F_\gamma= {\cal N}*\frac{\delta l}{d}$. This is the enhancement factor in the resonant case
\begin{equation}
F_\gamma= \frac{\delta l}{\lambda} \sqrt{\frac{c}{\lambda}}
\end{equation}
where $c$ is the speed of light.
The effective probability is then
\begin{equation}
{\cal P}_{\rm cham}(\omega)= F_\gamma P_{\rm cham}(\omega)
\end{equation}
The chameleon flux is
\begin{equation}
\Phi_{\rm cham}(\omega)= n_\gamma (R_{\rm res}) {\cal P}_{\rm cham}(\omega) p_\gamma (\omega)
\end{equation}
where $n_\gamma\sim 10^{21}\ {\rm s}^{-1}{\rm cm}^{-2}$ at the tachocline ($R= 0.7 R_\odot$) is the photon flux at the resonance and $p_\gamma$ the photon spectrum
\begin{equation}
p_\gamma(\omega)=\frac{\omega^2}{\pi^2\bar n} \frac{1}{e^{\frac{\omega}{T}}-1}
\end{equation}
where the average number of photons at temperature $T$ is
$
\bar n= \frac{2\zeta(3)}{\pi^2} T^3.
$
In the following we will assume that the photon flux is conserved and write
\begin{equation}
n_\gamma (r) = (\frac{0.7 R_\odot}{r})^2  10^{21} {\rm s^{-1}\cdot cm^{-2}}
\end{equation}
for the photon flux through a sphere of radius $r$.
\begin{figure}
\begin{center}
\includegraphics[scale=0.4]{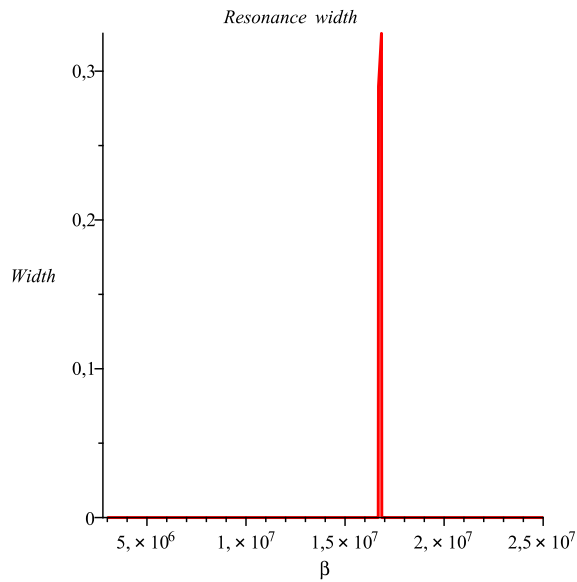}
\caption{The width of the resonance region in cm when $n=4$ as a function of $\beta=\frac{m_{\rm Pl}}{M}$ for $\beta_\gamma= 10^{10.32}$ and $\omega= 7 \ {\rm keV}$ when a  magnetic field is assumed to exist in a shell of width $0.01 R_\odot$ around the tachocline.  The width scales like $\beta_\gamma$ (see (\ref{wi})). Here we have chosen $\beta_\gamma$ which corresponds to the saturation of the solar energy loss bound by escaping chameleons. Notice that the width is of the order of  the photon mean free path $\approx 0.3$ cm here and much shorter than the width of the shell, i.e. the width of the shell does not influence the resonance. For each value of $\beta$ the location of the resonance can be deduced from Figure 2.}
\end{center}
\end{figure}

\subsection{Non-resonant production}

Chameleons can also be  created in the solar magnetic field when no resonance is present. In the non-resonant case, the conversion
probability in the sun is obtained from (\ref{prob})
\begin{equation}
P_{\rm cham}(\omega)=  4\theta^2 \sin^2\Delta
\end{equation}
as long as $\theta \ll 1 $ which is always true in the tachocline for a magnetic field of order 30T and photon energies less than say 2 keV.
On the other hand the behaviour of $\Delta$ inside the sun is highly dependent on the value of the coupling to matter $\beta$. For $\Delta\ll 1$, we can simply write  the production probability over a distance $l$ as
\begin{equation}
P_{\rm cham}(\omega)= \frac{B^2 l^2}{4 M_\gamma^2}
\end{equation}
For higher values  of $\Delta$, the production probability oscillates and we can write
\begin{equation}
P_{\rm cham}(\omega)= \frac{B^2 l_\omega ^2}{4 M_\gamma^2} \sin^2 \frac{l}{l_w}
\end{equation}
where we have introduced the coherence length
\begin{equation}
l_\omega= \frac{4\omega}{m^2_{\rm eff}}
\end{equation}
Another interesting phenomenon is that the production of chameleons is only allowed when
$\omega^2 \ge m^2_{\rm eff}$. This  becomes possible  in smaller and smaller regions around  the tachocline as $\beta$ increases. Eventually, when $\beta$ is large enough, no chameleons can be  produced in the sun.

Let us come back to the production mechanism in the sun when photons go through a magnetic region of size much larger than the photon mean free path. Let us consider all the possible photon paths in this region.
Photons perform a random walk with many changes of direction. Let us consider one such photon which covers a  radial distance $d(l)$ in one second. The number of scatterings along this trajectory is
\begin{equation}
N(l)=\frac{c}{l}
\end{equation}
where $l$ is the typical distance between two scatterings and
\begin{equation}
d(l)= l \sqrt{N(l)}
\end{equation}
The distance $l$ is random and distributed according to a Poisson law $e^{-\frac{l}{\lambda}} \frac{dl}{\lambda}$ corresponding to the absence of scattering over a radial length $l$. Notice that the average value of $l$ is given by the photon mean free path $<l>=\lambda$. We will see that, when the coherence length of photons $\l_\omega$ is small compared to the photon mean free path, the conversion probability is dominated by paths with scatterings after a length $l\sim l_\omega$. On the other hand, when $l_\omega$ is larger than the photon mean free path, the conversion probability is dominated by paths with $l\sim \lambda$.

Consider the radial region of size $\Delta R$ which contains the region of size $d(l)$ which is the radial distance covered by the photon in its random walk in one second and where both the mean free path length and the magnetic field remain unchanged. There are $\frac{\Delta R}{d(l)}$ such patches in a single region of size $\Delta R$ and $R_\odot/\Delta R$ regions over the entire sun, in principle. For a given length $l$, the probability of conversion in a region of size $\Delta R$ is
\begin{equation}
{\cal P}_{\rm cham} dl = \frac{\Delta R}{d(l)} N(l) P_{\rm cham}(\omega) e^{-\frac{l}{\lambda}} \frac{dl}{\lambda}
\end{equation}
where photons perform independent random walks with scatterings  following a  Poissonian distribution $e^{-\frac{l}{\lambda}} \frac{dl}{\lambda}$.
The total conversion probability in the sun is
\begin{equation}
{\cal P}_{\rm cham}= \sum_{i=1}^{\frac{R_\odot}{\Delta R}} \int_0^\infty {\cal P}_{\rm cham} dl
\end{equation}
When $\Delta R \ll R_\odot$ and $d\ll \Delta R$ as it is the case in reality, the conversion probability becomes
\begin{equation}
{\cal P}_{\rm cham}= \int_0^1 dx \int_0^\infty  \frac{ R_\odot}{d(l)} N(l) P_{\rm cham}(\omega,r) e^{-\frac{l}{\lambda(r)}} \frac{dl}{\lambda(r)}
\end{equation}
where $x=r/R_\odot$ is the rescaled radial distance and
\begin{equation}
{ P}_{\rm cham}(\omega,r)= \frac{B^2 l_\omega(r) ^2}{4 M_\gamma^2} \sin^2 \frac{l}{l_w(r)}
\end{equation}
with\begin{equation}
l_\omega(r)= \frac{4\omega}{m^2_{\rm eff}(r)}
\end{equation}
and  the effective mass depends on $r$. Typically, we will find that this conversion probability can be as large as $10^{-9}$.
Using $y=l/l_\omega$ we find that
\begin{equation}
\frac{d{\cal P}_{\rm cham}}{dx}= \sqrt{\frac{c}{l_\omega(r)}} \frac{B^2(r)l_\omega^2(r)R_\odot }{4 M_\gamma^2 \lambda (r)} \int_0^\infty \frac{\sin^2(y)}{y^{3/2}} e^{-y \frac{l_\omega(r)}{\lambda(r)}} dy
\end{equation}
When $l_\omega$  is much larger than the mean free path, the probability is dominated by paths of order $l\sim \lambda$ and
\begin{equation}
\frac{d{\cal P}_{\rm cham}}{dx}= \sqrt{\frac{c}{l_\omega(r)}} \frac{B^2(r)l_\omega^2(r)R_\odot}{4 M_\gamma^2 \lambda (r)} \int_0^{\lambda(r)/l_\omega (r)} y^{1/2} dy= \sqrt{\frac{c}{\lambda (r)}} \frac{B^2(r)\lambda (r)R_\odot }{24 M_\gamma^2 }
\end{equation}
When the coherence length $l_\omega$ is smaller than the mean free path, $l_\omega \ll \lambda$, the conversion probability  is given by
\begin{equation}
\frac{d{\cal P}_{\rm cham}}{dx}=\sqrt{\frac{c}{l_\omega(r)}} \frac{B^2(r)l_\omega^2(r) R_\odot}{4 M_\gamma^2 \lambda (r)} \int_0^\infty \frac{\sin^2(y)}{y^{3/2}} dy
\label{prr}
\end{equation}
Of course this is only realistic if the coherence length is large enough that the presence of the solar plasma can be neglected. Typically, we will find cases where the coherence length can be as low as $10^{-2}\lambda$ ($\lambda$= mean free path of the thermal photons). In the following we will cut the probability distribution below the scale  $\lambda_{\rm pl}=\omega_{\rm pl}^{-1}$. When the coherence length drops below the plasma scale $l_\omega \lesssim \lambda_{\rm pl}$, we average over the oscillatory behaviour of the conversion probability over a length of the order of the mean free path and  we get
\begin{equation}
\frac{d{\cal P}_{\rm cham}}{dx}= \sqrt{\frac{c}{l_\omega(r)}} \frac{B^2(r)l_\omega^2(r)R_\odot }{4 M_\gamma^2 \lambda (r)} \int_{\lambda/l_\omega}^\infty \frac{1}{2y^{3/2}} dy\sim \sqrt{\frac{c}{\lambda(r)}} \frac{B^2(r)l_\omega^2(r)R_\odot }{4 M_\gamma^2 \lambda (r)}
\label{pro}
\end{equation}
The differential flux of emitted chameleons is then
\begin{equation}
\frac{d\Phi_{\rm cham}}{dx}= n_\gamma p_\gamma \frac{d{\cal P}_{\rm cham}}{dx}
\end{equation}
This flux of chameleons escape from the sun and reaches the  earth where they can be back-converted to photons inside a magnetic helioscope like CAST and Sumico. A small fraction of these chameleons can also be back-converted in the magnetised photosphere and above.

A strong constraint on chameleons arises from the requirement that the energy carried away by chameleons does not affect the life time of the sun. In practice, we impose that chameleons cannot carry more than 10 percent of the total luminosity of the sun,  estimated to be $6\cdot 10^{10}\ {\rm erg\cdot s^{-1} \cdot cm^{-2}}$ \cite{Gondolo:2008dd}, which then would not affect the evolution of the sun over its lifetime.
The total luminosity loss in chameleons is given by
\begin{equation}
L_{\rm cham}= \int d\omega \omega \Phi_{\rm cham}( \omega)
\end{equation}
where the chameleon flux must include both the resonant and non-resonant contributions.
In the numerical examples which will be presented, the energy loss bound is satisfied.

\subsection{Back-conversion in a helioscope}

Let us consider a magnetic helioscope like CAST at the surface of the earth. Some of the incoming chameleons emerging from the sun will be back-converted provided their energy is higher than their mass in the atmosphere and in matter in front of the helioscope entrance. Let us first focus on the  resonant case. The number of back-converted photons is
\begin{equation}
\frac{dN_\gamma}{d\omega}= 4\frac{A R_{\rm res}^2}{D^2} \Phi_{\rm cham} (\omega) \theta^2_{\rm helio} \sin^2 \Delta_{\rm helio}
\end{equation}
where D is the earth-sun distance (1 AU), $A$ the aperture  and the conversion probability
\begin{equation}
P_{\rm helio}=4\theta^2_{\rm helio} \sin^2 \Delta_{\rm helio}
\end{equation}
simplifies when the coherence length in the helioscope is larger than the magnetic pipe length
\begin{equation}
P_{\rm helio}=
\frac{B^2_{\rm helio} L^2}{4 M_\gamma^2}
\end{equation}
where $L$ is the helioscope pipe length.
Integrating over the energy,  this gives the total number of  back-converted photons from solar chameleons entering the helioscope pipe:
\begin{equation}
N_\gamma = N_0 \beta_\gamma^7
\end{equation}
 and
\begin{equation}
N_0= \frac{ f(5)}{8 \zeta(3)} F_\gamma \frac{R^2_{\rm res} A}{D^2} \frac{n_\gamma B^2_{\rm helio} L^2 T^2_{\rm res} B^5_{\rm res}}{m_{\rm Pl}^7} (\frac{dl}{d\rho}\vert_{\rm res})^3
 (\frac{(1+n) m_e m_p}{4\pi \alpha})^3
\end{equation}
with $\zeta(.)$ the Riemann $\zeta$ function and  $T_{\rm res}$ the solar temperature at  the resonance, furthermore:
\begin{equation}
f(5)= \int_{\frac {\omega_{\rm min}}{T_{\rm res}}}^{\frac {\omega_{\rm max}}{T_{\rm res}}} dx \frac{x^5}{e^x -1}
\end{equation}
where the bounds of the integral depend on the detector energy range . In the following we will take $\omega_{\rm min}=1 \ {\rm keV}$ and $\omega_{\rm max}=7 \ {\rm keV}$
corresponding to the present CAST detectors.

\begin{figure}
\begin{center}
\includegraphics[scale=0.4]{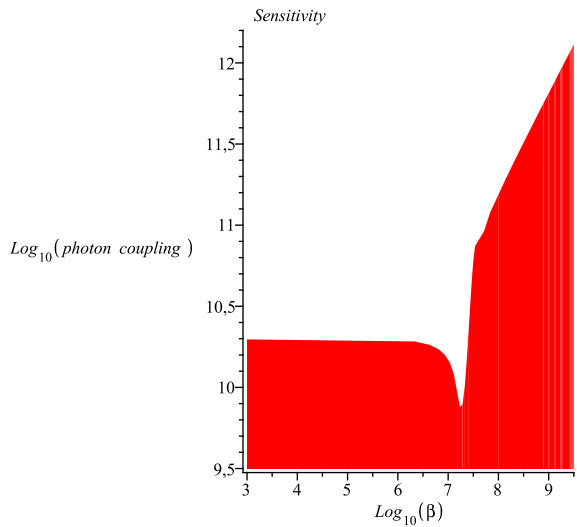}

\caption{The sensitivity region at the 2$\sigma$ level above the red region for a magnetic helioscope like CAST in vacuum  at a temperature of 1.8K and for the photon coupling $\log\beta_\gamma$ as a function of the matter coupling $\log\beta$. We have assumed that there is a constant magnetic field of 30T in a shell of $0.01 R_\odot$ surrounding the tachocline. The precise form of the diagram depends on $n$ although the overall shape remains the same. Notice that for values of $\beta$ below the resonance values, where the dip occurs, the 2$\sigma$ contour is constant. In this work, we have chosen this value as a template for the photon coupling. Larger values of the photon couplings would violate the solar energy loss bound due to  escaping chameleons with $\beta\lesssim 10^{10.32}$. As such, the solar model we have chosen for the magnetised region, where chameleons are produced, is almost optimal. Notice that the value $\beta=\beta_m$ would only lead to interesting physics at the 2$\sigma$ level for very large values of the coupling to photons, higher than the CHASE bound.
}
\end{center}
\end{figure}

In the non-resonant case (section III-B),
the  number of regenerated photons in the helioscope pipe emitted from a non-resonant region is obtained by multiplying the production flux by the conversion probability in the helioscope pipe:
\begin{equation}
\frac{d^2N_\gamma}{d\omega dx}= 4\frac{A x^2 R_\odot^2}{D^2} \theta^2_{\rm helio} \sin^2 \Delta_{\rm helio}  \frac{d\Phi_{\rm cham}}{dx}
\end{equation}
When the coherence length  in the helioscope pipe is larger than $L$, we find that
\begin{equation}
\frac{d^2N_\gamma}{d\omega dx}= \frac{A x^2 R_\odot^2}{D^2}  \frac{B^2_{\rm helio} L^2}{4 M^2_\gamma} \frac{d\Phi_{\rm cham}}{dx}
\end{equation}
and upon integrating over the sun radial distance $x$, we find that
\begin{equation}
\frac{dN_\gamma}{d\omega}= \frac{A  R_\odot^2}{D^2}  \frac{B^2_{\rm helio} L^2}{4 M^2_\gamma} \int_0^1 dx\ x^2 \frac{d\Phi_{\rm cham}}{dx}
\end{equation}
The overall dependence on the photon coupling is $\beta_\gamma^4$ as $\frac{d\Phi_{\rm cham}}{dx}$ varies with  $\beta_\gamma^2= \frac{m^2_{\rm Pl}}{M_\gamma^2}$. The dependence on the matter coupling enters essentially in the effective mass inside the sun and therefore in the coherence length in the sun.
\begin{figure}
\begin{center}
\includegraphics[scale=0.4]{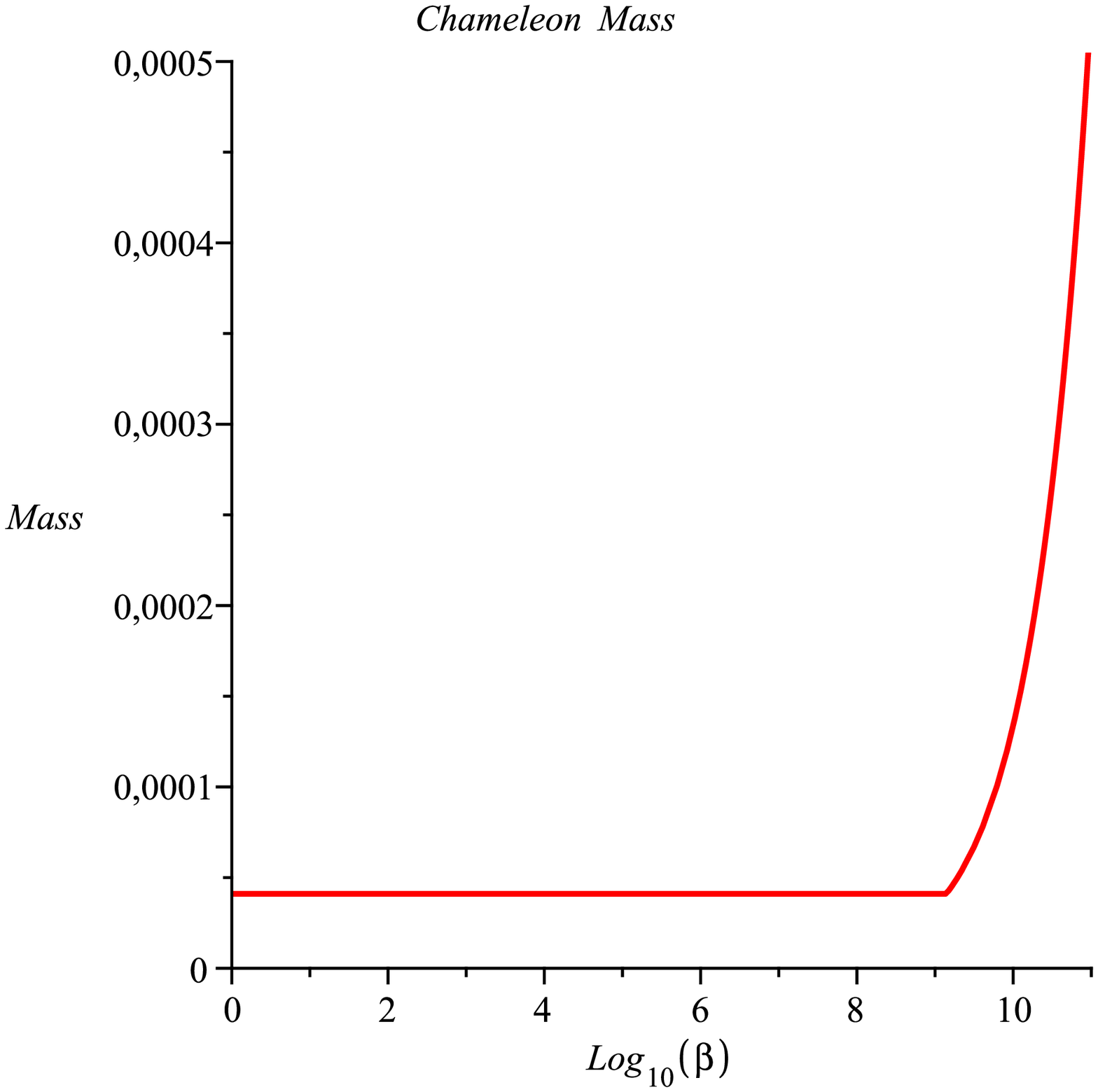}

\caption{The mass of the chameleon  in eV in the helioscope pipe  (vacuum) as a function of $\log \beta$. The mass is constant and independent for low values of $\beta\lesssim 10^9$, and becomes $\beta$ dependent for larger values. The constant value is obtained from (\ref{ma}) and the $\beta$ dependence from (\ref{mass}).}
\end{center}
\end{figure}

As a summary, we have found that a resonance can always generate chameleons for values of the matter coupling $\beta$ in a narrow range for a given value
of the index $n$. On the other hand, non-resonance production is generic and happens for the large range of allowed values $\beta\lesssim 10^{11}$. It should also be noticed that for values of $\beta$ larger than a $n$-dependent value, no chameleon can be produced in the sun. This value is typically around $\beta=10^{11}$ for $n=4$ implying that chameleons can always be non-resonantly produces for $n=4$ and $\beta\lesssim 10^{11}$. In the following, we will concentrate on a single value of $\beta=10^6$ where no resonance is present and analyse the properties of the non-resonant production.

\begin{figure}
\begin{center}
\includegraphics[scale=0.4]{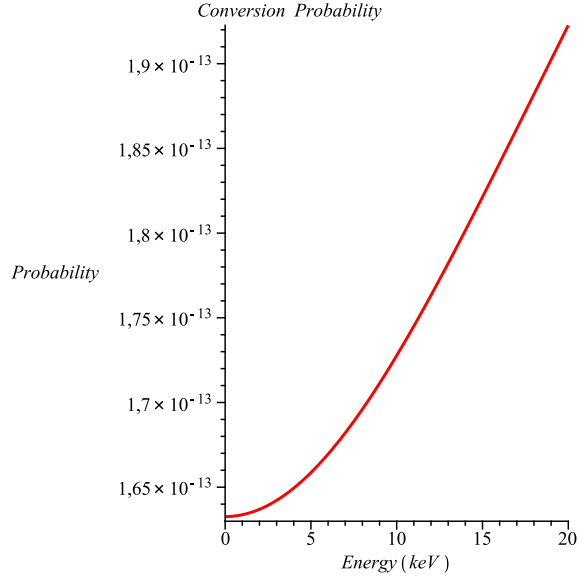}

\caption{The conversion probability  as a function of the photon/chameleon energy in keV with vacuum  in the magnetic helioscope pipe. Here we have chosen the matter coupling to be $\beta=10^{6}$ and the photon coupling $\beta_\gamma= 10^{10.32}$. The magnetised region is such that $B_{\rm helio} L\approx 90\ T\cdot m$. Notice the parabolic shape at low energy followed by a sinusoidal trend which would saturate at much  higher energies (not shown here).
}
\end{center}
\end{figure}

\subsection{Back-conversion in the photosphere and above}

Before leaving the sun, a tiny fraction of the produced chameleons can be back-converted into photons while propagating through the magnetised photosphere and the other layers of the solar atmosphere.
We will be interested in three  types of conversion processes. The first one would happen in the photosphere below the solar surface. Inside the sun,  the mean free path of photons is the maximal  extension over which a back-conversion can occur. Indeed, on larger distances, photons are scattered and cannot mix with the incoming chameleonic wave function. In the photosphere, the magnetic field is of order 0.2 T. Above the solar surface, we envisage two situations. The first one corresponds to the quiet sun with a small magnetic field of order 1 Gauss distributed over large regions satisfying the empirical law\cite{BL}
\begin{equation}
B_{\rm out} L_{\rm out} \sim 10^4 \ T\cdot m
\label{emp}
\end{equation}
We also consider flares where the magnetic fields are much larger while the size of the magnetic regions is smaller although still  following the same empirical law (\ref{emp}).

The number of back-converted  photons is given by
\begin{equation}
\frac{dN_{\rm out}}{d\omega}= \frac{B^2_{\rm out}{L_{\rm out}^2}}{4M^2_\gamma} \Phi_{\rm cham}
\end{equation}
when  the coherence length (in the photosphere) is much larger than the size of the magnetic region. In the photosphere where the density is relatively high compared to the atmosphere, the coherence length is relatively short and one must average out over the distribution of photon excursions between two scatterings. This implies, using (\ref{prob}) and
the averaged value $<\sin ^2 \Delta_{\rm out}>=\frac{1}{2}$, that the number of back-converted photons is
\begin{equation}
\frac{dN_{\rm out}}{d\omega}= 2 \theta_{\rm out}^2  \Phi_{\rm cham}
\end{equation}
where
\begin{equation}
\theta_{\rm out} = \frac{\omega B_{\rm out}}{ M_\gamma m^2_{\rm eff}}
\end{equation}
is reduced by the larger effective mass in denser regions, such as inside the photosphere compared to the corona. Indeed, we find that the back-conversion of outstreaming chameleons from the inner sun  is largely suppressed in the photosphere where the density is higher than in the upper atmosphere with a much smaller plasma density. As a result we find that the back-conversion of photons from outstreaming chameleons is mainly taking place  in magnetised regions above the chromosphere.

As the back-converted photons have a spectrum in the soft X-ray band, this leads  to an excess of solar X-ray luminosity  given by
\begin{equation}
L_{\rm out}=\int_0^\infty d\omega \omega \frac{dN_{\rm out}}{d\omega}
\end{equation}
It is shown below that this radiation excess is tiny compared to the total  solar luminosity. However,   the solar luminosity in the  soft X-ray band due to such back-converted photons for the quiet sun or the active sun is of interest  as it could explain the unaccounted part of the solar photon spectrum (solar corona problem).

\begin{figure}
\begin{center}
\includegraphics[scale=0.4]{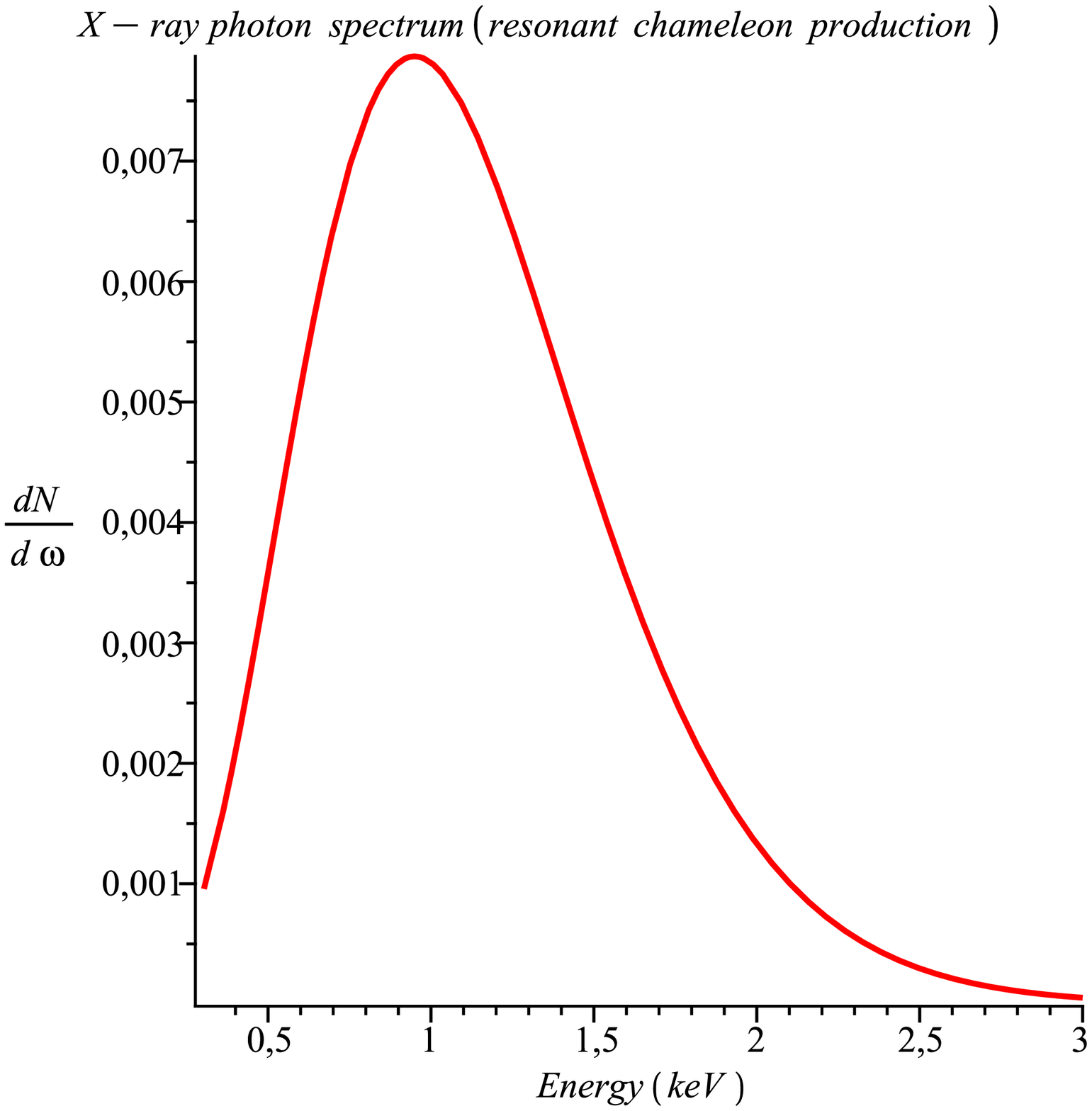} \hspace*{1cm}
\includegraphics[scale=0.4]{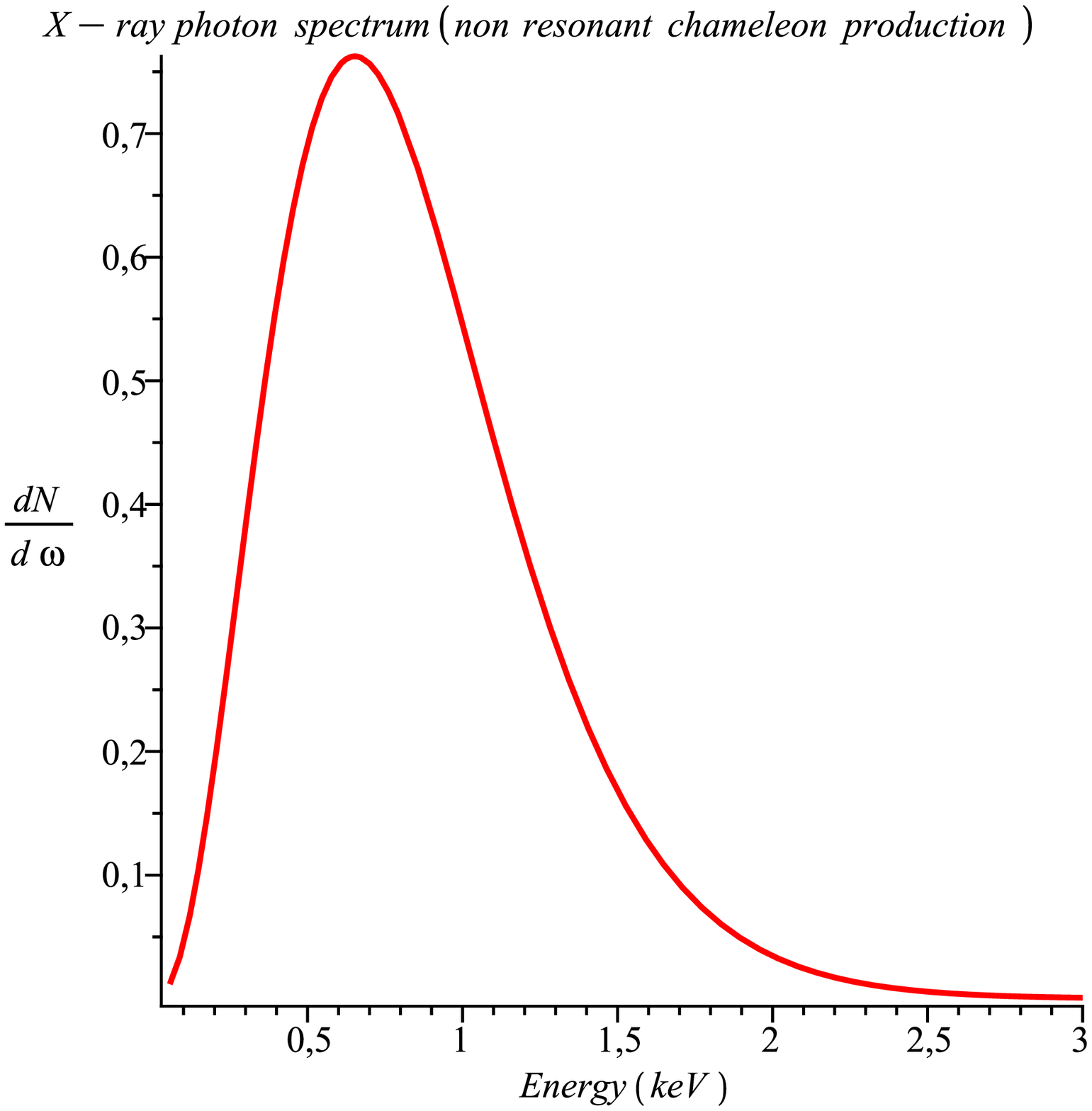}
\caption{The resonant spectrum of back-converted photons from solar chameleons giving the  number of counts per hour and per keV
predicted to be seen in   a magnetic helioscope like CAST in vacuum     as a function of the reconverted photon energy.  Here we have chosen the matter coupling to be $\beta=1.8\cdot 10^{7}$, the photon coupling $\beta_{\gamma}=10^{10.29 }$ and a shell magnetic field in the sun of 30T over a width of 0.01 solar radius above the tachocline for the resonant case (left) and   for the non resonant case (right) with $\beta=10^6$ and $\beta_\gamma=10^{10.29}$. }
\end{center}
\end{figure}

It is therefore relevant  to give  an order of magnitude estimate of the predicted X-ray intensity from back-converted solar chameleons and compare it with observation. Since the chameleon  effective mass is low in sparse regions, it decreases outside the sun  when chameleons propagate in empty space. Moreover the coherence  length  $l_\omega$ increases with $1/m_{eff}^2$ and the conversion probability with $B_{\rm out}^2 L_{\rm out}^2$. In the weak but extended magnetic fields $B_{\rm out}$ which is present all the way to the vicinity of the Earth and in particular the orbiting X-ray observatories, the chameleon-to-photon conversion is possible. As an illustration,  in Figure 6 we see that  the chameleon-to-photon  conversion efficiency is greater than $1.65\cdot$ 10$^{-13}$ in a CAST-like pipe for BL=90 Tm. The estimated integral of the inter planetary magnetic field $B_{\rm out} L_{\rm out}$  is about $B_{\rm out} \cdot L_{\rm out} \approx 10^{3}$ to $10^{5}$Tm and occasionally even larger \cite{zpatras}. The conversion probability becomes then 100 times to $10^6$ times larger than the one estimated for a CAST-like  magnet in vacuum (see Figure 6). Taking into account that the maximal  allowed solar luminosity in any invisible channel is about 10\% of the total solar luminosity, then, a rather conservative estimate of the expected excess X-ray observation at 1 AU, is about $6\cdot 10^{21}$ to  $6\cdot 10^{25}$erg/s.  Surprisingly, this fits observations (see ref. \cite{zpatras} and references therein).

How could  the chameleon origin of this excess  be established observationally in a non-equivocal manner? In fact, the following two measurements can be performed: 1) the measurement of the  solar X-ray ray spectrum in the energy range below $\sim$ 2 keV, which has been done in the recent solar minimum \cite{zpatras}, and 2) the two dimensional reconstruction of the quiet sun X-ray spectrum in the same energy range. If the tachocline were the source of the outstreaming chameleons which would back-convert to soft X-rays on their journey to Earth through the  magnetic field (only its transverse component matters), the derived two dimensional plot would have a characteristic ring-like structure at about 0.7$\cdot R_{\odot}$. This would be a clear signature of the creation of chameleons or other similar exotica  around the tachocline. Further studies, following the suggestions of this work, would allow to localize the involved processes.
Moreover, once reconstructed, this would provide a new tool for investigating the solar plasma, either in the interplanetary space or deep inside the Sun, i.e. near the tachocline some 200000 km deep inside the Sun. This would have the potential to become, along with the solar neutrinos and helioseismology, one of  the most precise  methods to obtain information from the otherwise perfectly shielded inner Sun.

Of course, laboratory measurements as suggested here (see section IV), or, observations with a magnetic helioscope like CAST would provide complimentary information. In fact, CAST can increase its sensitivity to solar chameleons by a factor of 10-100 in solar chameleon signature, once its detectors can perform with a lower energy threshold, i.e.  towards a  few 100 eV's (see also \cite{zspsc}).

\section{Chameleon through Wall Experiment}

So far we have presented mechanisms for the creation of chameleons in the sun by the Primakoff effect and their back conversion at the surface of the magnetised sun or in magnetic helioscopes by the inverse Primakoff effect. Of course these results depend on the choice of the solar model for the magnetic region where chameleons are produced. It would be extremely important to have a model-independent probe of the existence of chameleons in the X-ray domain. This would also complement the CHASE experiment which has been carried out in the eV energy range and might suffer from uncertainties related to multiple reflections of chameleons at the beamtube walls.

We suggest
an experiment where a powerful source of X-rays shines through a , say,  CAST-like magnetic  pipe,  while another identical pipe lies downstream on the
other side of a X-ray absorbing thick  barrier. Of course, on the other side of the barrier no photons should appear conventionally. Now the possibility of creating energetic chameleons in the first magnet which would go downstream through the thick  barrier allows the regeneration of photons on the other side of the barrier, that by the time-reversed process.  We consider that both magnetic pipes on each side are identical.  In practice we can have  $n_\gamma\approx  10^{19} /\rm cm^2\cdot s$ at most  so far.
The spectrum of regenerated photons is
\begin{equation}
\frac{d{\cal N}_{\rm wall}}{d\omega}= 16 A \theta_{\rm helio}^4  \sin^4 \Delta_{\rm helio} P_{\rm source} (\omega)
\end{equation}
where $\theta_{\rm helio}\ll 1$ in the helioscope pipe. We have taken the source spectrum to be $P_{\rm source}$ such that
\begin{equation}
n_\gamma= \int_0^\infty d\omega \ P_{\rm source} (\omega)
\end{equation}
and $A$ is the aperture of the pipe.  Upon integration this spectrum, we obtain the total number of photons per second
\begin{equation}
{\cal N}_{\rm wall}= \int_0^\infty  d\omega \ 16 A \theta_{\rm helio}^4  \sin^4 \Delta_{\rm helio} P_{\rm source} (\omega)
\end{equation}
When this number is higher or at least not much  smaller  than the background level noise of the detectors in the appropriate energy range, this gives a new way of searching for chameleons, independently of any astrophysical modeling.

Of course axions with a similar mass and photon coupling as chameleons would also go through the dense barrier and regenerate X-rays with a resulting spectrum very similar to the chameleonic case. So how can one distinguish chameleons from axions? In fact, one can use one of the fundamental properties of chameleons: the dependence of their mass on the density in the pipe, while QCD-inspired axions have a fixed rest mass. Indeed as we have already exemplified in figure 8, when the density of the gas filling the pipe is increased, the effective mass of the chameleons becomes also larger and the conversion probability oscillates and is suppressed by $1/m^2_{\rm eff}$, hence a quartic signal suppression in the chameleon through wall experiment. As a result, for a high enough pressure, the number of regenerated photons would become essentially zero. This is a distinguishing property of chameleons. Indeed, one could envisage to perform the experiment in vacuum and then with an optically transparent gas in the pipe at a relatively high pressure,
exploiting the different dependencies of the chameleon and ALP production on the gas pressure in the beam pipe. In fact, a CAST pressure of $\approx $ 1 mbar is transparent to X-rays and should sufficient to realise the identification of chameleons.  A similar technique will be applied for the detection of the chameleon force between metallic plates: by comparing the force in vacuum and with a dense material between the plates, one may detect the presence of chameleons. Here the same type of technique may give a direct indication of the existence of chameleons in the keV energy range.

\section{Results}

\subsection{Solar chameleons}

For a given $n$, the resonance inside the sun depends on $\beta$. We find that values of $\beta$ which lead to a resonance are quite limited. Outside this small interval, no resonance is present. In figure 1, we have plotted the density at the resonance and in figure 2 the solar radius of the resonance as a function of $\beta$ for $n=1$ and $n=4$. Typically the resonance exists for $6\cdot 10^6\le \beta \le 4\cdot 10^7$ for $n=4$. For large values, $\beta\gtrsim 10^{11}$, no chameleon can be created inside the sun when we focus on a production region of width 0.01 solar radius around the tachocline at $0.7\ R_\odot$. Larger values of $\beta$ are already excluded by nuclear physics measurements of the neutron energy levels in the terrestrial gravitational field\cite{Brax:2011hb}. Different values of $n$ give almost the same type of results.

When it comes to the back-conversion of the emitted chameleons in a helioscope, we concentrate on  CAST-like conditions.
We take into account  100 hours of observation  with a detection noise of 1 photons per hour in the 1-7 keV range when the vacuum  in the pipe is such that the pressure is about $10^{-8}$ mbar.  We have taken the pipe length to be 9.26 m, the magnetic field  9~T and the diameter of the pipe to be 43~ mm. In the sun, we have  considered  a constant magnetic field of 30 T above the tachocline in a shell of width $0.01 R_\odot$. In this case, the width of the resonance as a function of $\beta$ is given in Fig. 3. It is of the same order of magnitude as the thermal photon mean free path in the tachocline, i.e. around 0.3 cm.
When observations are carried out during h hours, the signal to noise ratio is
\begin{equation}
\sigma= \frac{hN_\gamma}{\sqrt{h(N_\gamma + N_{\rm helio})}}
\end{equation}
where the noise level of the helioscope is taken to be $N_{\rm helio}=1$ per hour as a generic example. We have given in Fig.4 a sensitivity plot at the 2$\sigma$ level for a 100 hour measurement in vacuum. As can be seen, for values of $\beta$ below the resonance range, the sensitivity region is uniform and the matter coupling must be
\begin{equation}
\beta_\gamma \ge 10^{10.29}
\label{bound}
\end{equation}
which is to be compared with  the CHASE bound $\beta_\gamma\le 10^{11}$.
Outside the resonance values, the coherence length in the sun decreases rapidly and the production probability oscillates very fast leading to a diminishing sensitivity.

We note that inside a helioscope  pipe in  vacuum, the rest mass of the chameleon is determined by the radius $R_{\rm helio}$ of the pipe and is density independent\cite{optics}
\begin{equation}
m_{\rm vac}= \frac{2\sqrt{n+1}}{R_{\rm helio}}
\label{ma}
\end{equation}
This is valid at low values of $\beta$. For higher values, the density dependent rest mass becomes the chameleon mass (\ref{mass}), see figure 5.

\begin{table}[]

\begin{tabular}{|c|c|c|c|}
\hline
\hline ${\rm Altitude \rm{(km)}} $ & $\rho (\rm {g\cdot cm^{-3}})$ & $\lambda \rm{(km)}$ & $B (\rm{Gauss})$  \\
\hline -500 & $7.4\ 10^{-7}$ & 15 & 2000  \\
\hline -100 & $2.74\ 10^{-7}$& 40 & 2000 \\
\hline 0 & $2\ 10^{-7}$& 50 & 2000 \\
\hline 500 & $6.79 \ 10^{-9}$& $10^4$ & 1000 \\
\hline 1000 & $10^{-10}$ & $10^5$ & 100 \\
\hline 2000 & $3.4 \ 10^{-13}$ & $10^8$ & 10 \\
\hline 10000 & $10^{-15}$ & $10^{10}$ & 10 \\
\hline 80000 & $2 \ 10^{-14}$ & $10^9$ & 5 \\
\hline
\end{tabular}
\caption[]{The values of the density, photon mean free path and magnetic field as a function of the solar altitude from the solar surface. Above an altitude of 1000 km, the considered cases correspond to dynamically changing solar flares.}
\end{table}

We have studied the resonant and non-resonant spectra obtained after back-conversion in the helioscope pipe using for the photon coupling $\beta_\gamma=10^{10.29}$ and matter couplings ($\beta=10^6$, $\beta= 1.8\cdot 10^7$) in the non-resonant and  the resonant case respectively. As can be seen in Fig. 7, the number of resonantly back-converted photons is lower than the non-resonant ones. The width of the resonant region in $\beta$ is so narrow (see figure 3) that one may ignore the resonant case and therefore we focus on the non-resonant one which can occur for all values of $\beta\lesssim 10^{11}$. An important consequence of the non-resonant production is that  the resulting X-ray spectra appear in  sub-keV energies with a peak around  $\approx$ 600 eV. This is  below the present sensitivity of all the magnetic helioscope X-ray detectors. With a better sensitivity in a lower energy range, the potential to detect chameleons would be enormously enhanced. This is an exciting and challenging experimental possibility.

We have used the present sensitivity levels of magnetised helioscopes to deduce that values of the coupling constant to photons $\beta_\gamma= {\cal O}(10^{10.29})$ should
be within experimental reach. With an improved sensitivity for detectors in the sub-keV region, we expect that lower values of the photon coupling constants should
lead to a detectable number of detected X-ray photons. Of course, if the coupling to photons $\beta_\gamma$ is extremely low, it is unlikely that any signal could be detected. The analysis of the range of $\beta_\gamma$ with sub-keV detectors is left for future work.

We have also investigated what happens when the pressure in the helioscope pipe is not vacuum. As shown in figure 8, the number of photons decreases as the pressure increases. This is due to the fact that the chameleon mass increases with the pressure. When the coherence length is smaller than the effective length of the helioscope, the conversion probability decreases as $1/m^2_{\rm eff}$ implying a decrease of the overall amplitude. Moreover, rapid oscillations are present for larger pressures. This is the dependence that we suggest to utilise for (solar) chameleon identification.

We are also interested in the chameleon spectrum on earth. The number of chameleons per second and $\rm cm^2$ has been plotted in Fig. 9.
A crucial test of our  model is whether the luminosity carried by the chameleons is below the energy loss bound  for any unknown type of escaping solar radiation which is  10 \% of the total solar luminosity\cite{Gondolo:2008dd}. In figure 9, we have plotted the analogous spectrum of the escaping chameleons which  corresponds to  $\lesssim$ 10 \% of the total solar luminosity. The saturating bound is
\begin{equation}
\beta_\gamma \le 10^{10.32}
\end{equation}
As can be seen, the solar model with a magnetised region of width  $0.01 R_\odot$ at the tachocline and a magnetic field of 30T is quite realistic.
\begin{figure}
\begin{center}
\includegraphics[width=7cm]{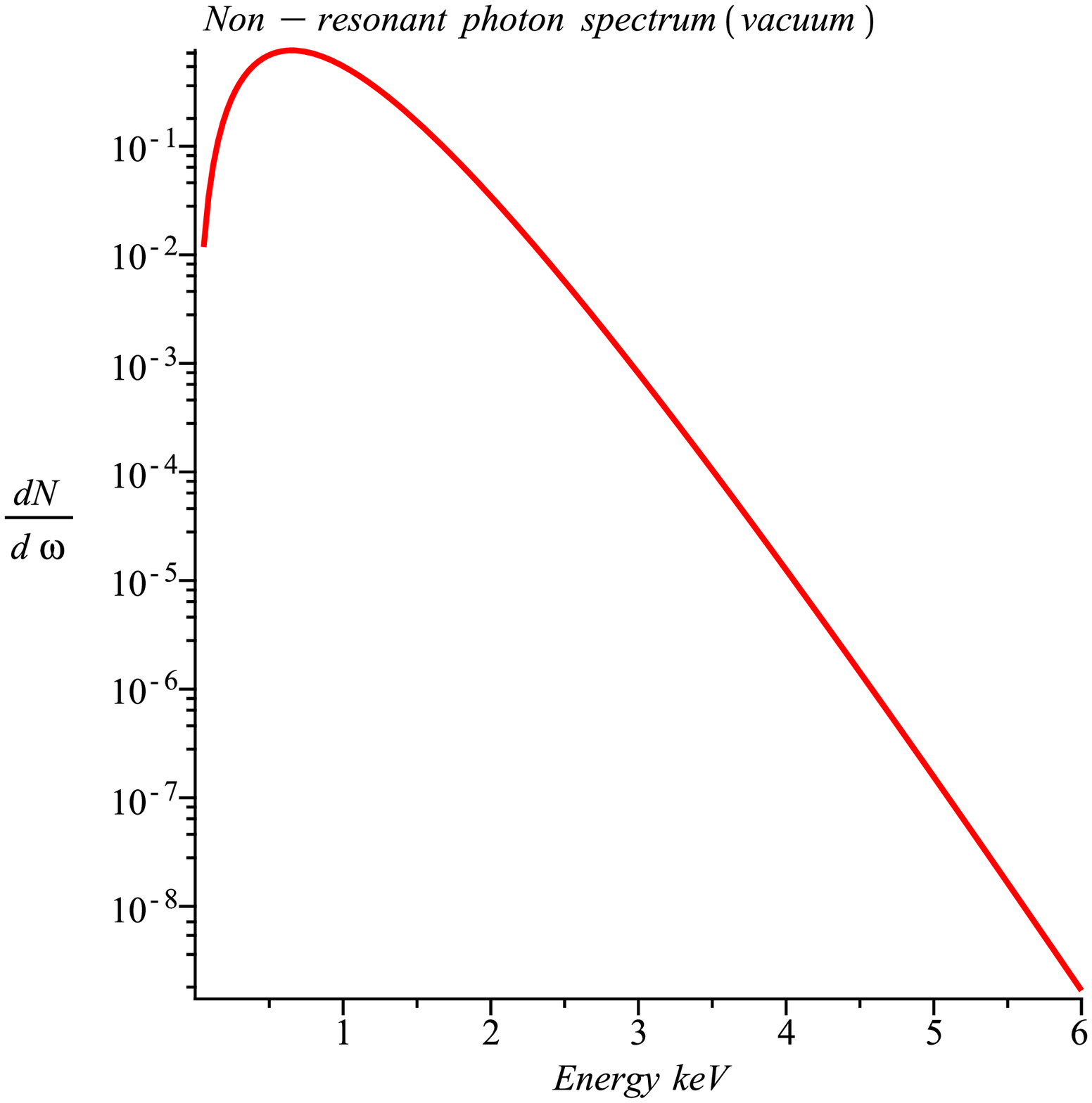} \hspace*{1cm}
\includegraphics[width=7cm]{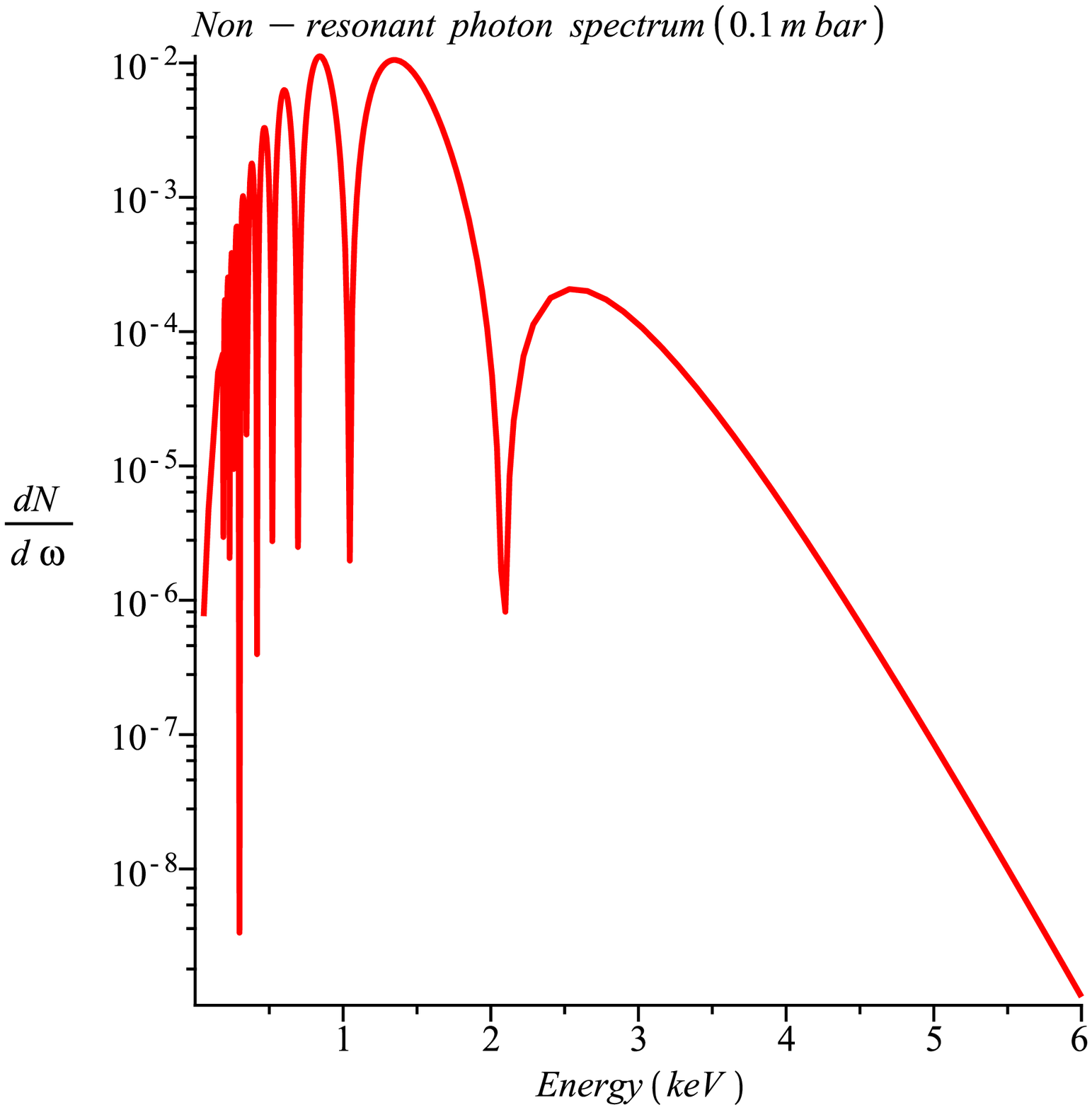} \vspace*{1cm}\\
\includegraphics[width=7cm]{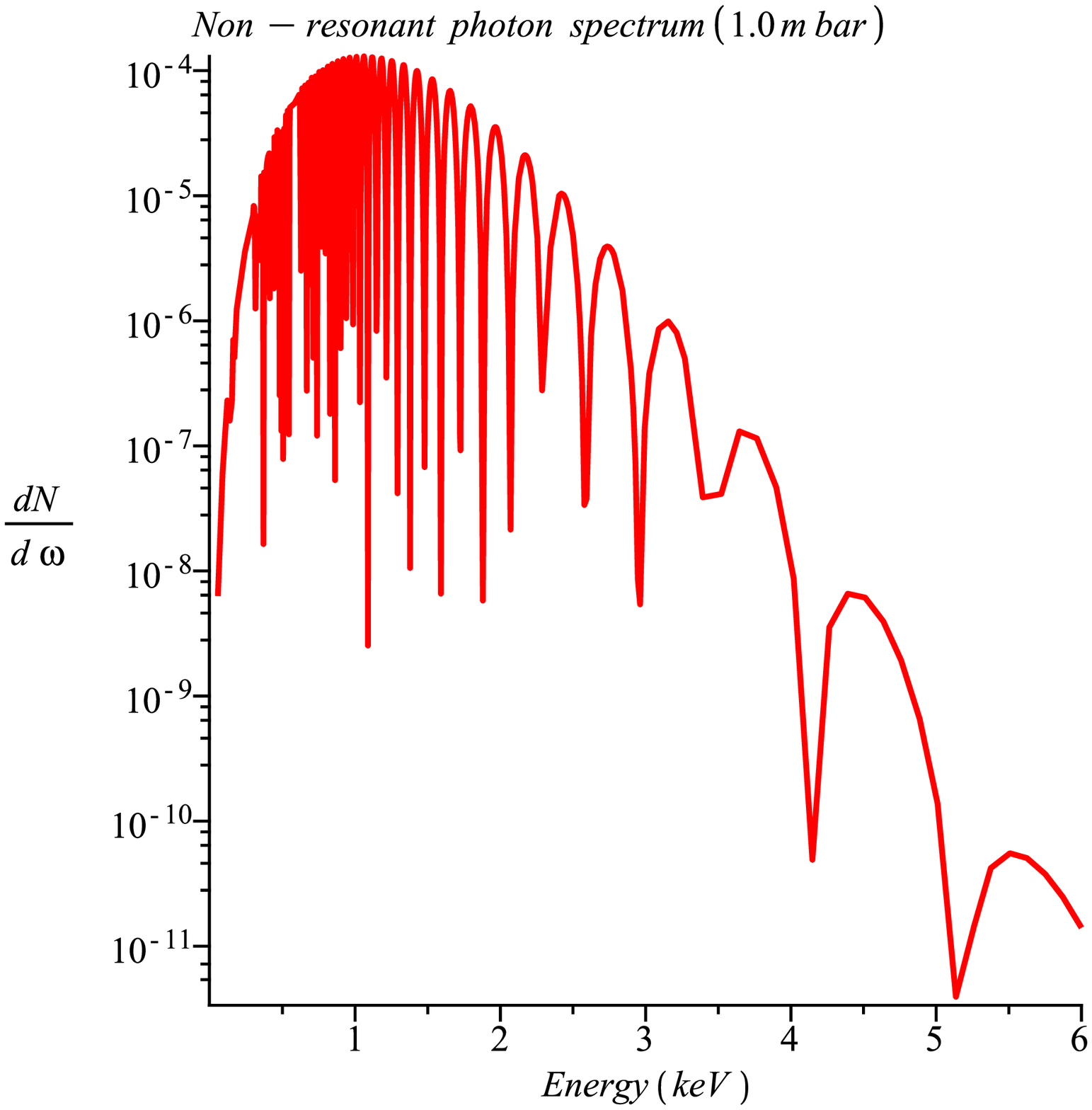} \hspace*{1cm}
\includegraphics[width=7cm]{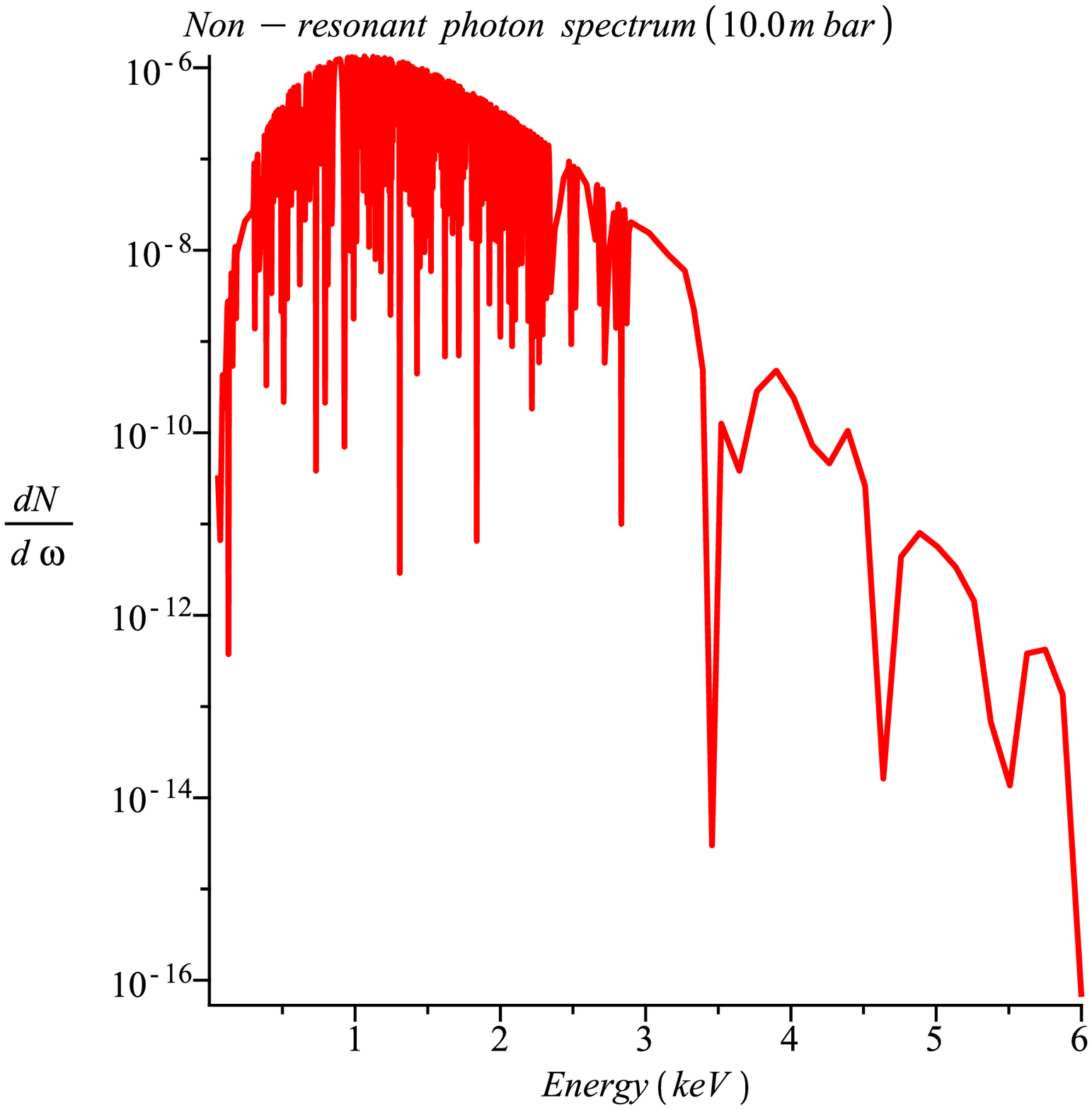}
\caption{The spectrum of regenerated photons giving the  number of counts per hour and per keV
as predicted to be seen by   a helioscope  like CAST  as a function of the photon energy in keV.
Here we have chosen the matter coupling to be $\beta=10^{6}$, the photon coupling $\beta_{\gamma}=10^{10.29}$ and a shell magnetic field in the sun of 30T over a width of 0.01 solar radius near the tachocline.}
\label{SpecExp}
\end{center}
\end{figure}

\begin{figure}
\begin{center}
\includegraphics[width=7cm]{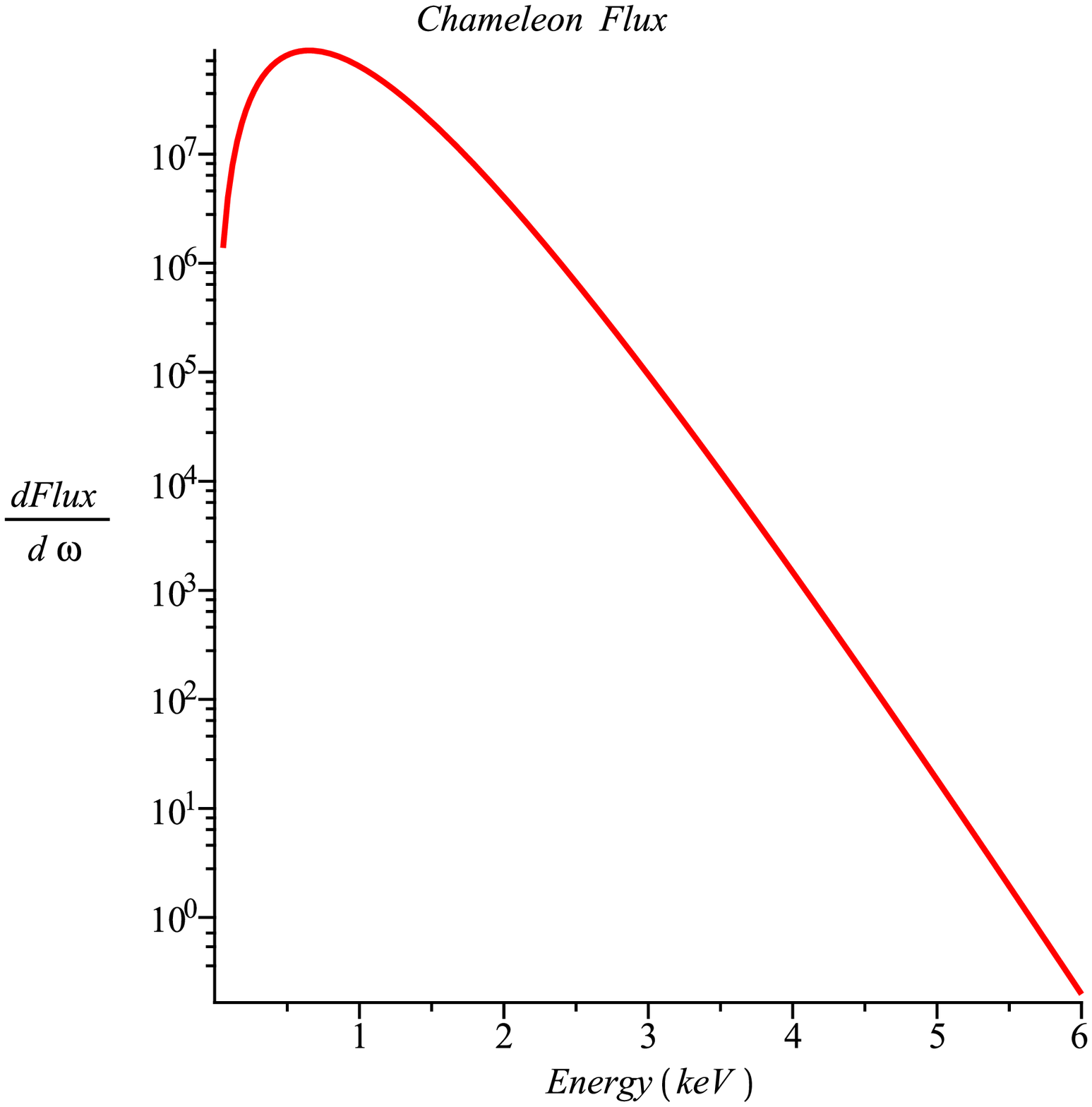}\hspace*{1cm}
\includegraphics[width=7cm]{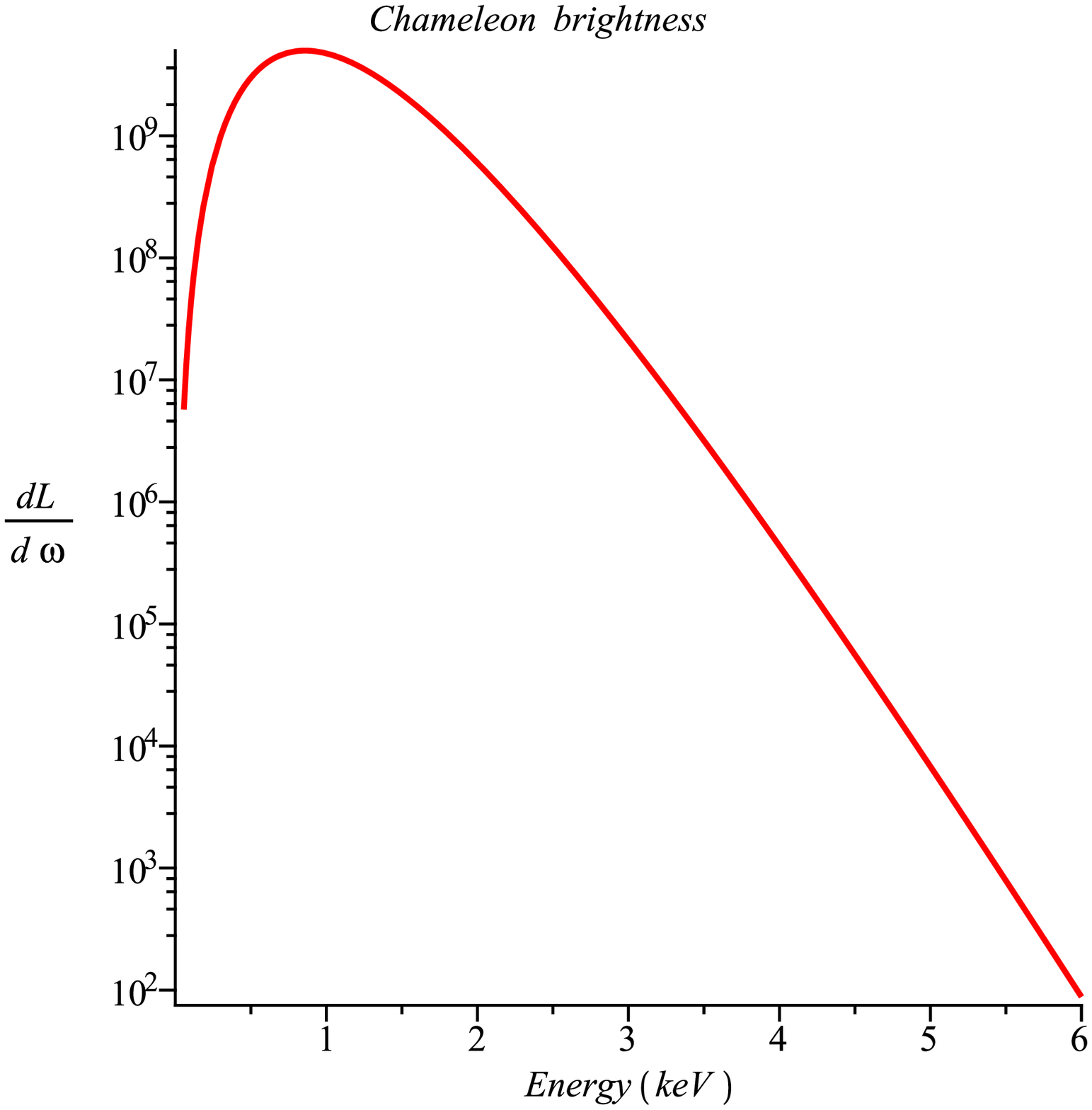}\vspace*{1cm}\\
\includegraphics[width=7cm]{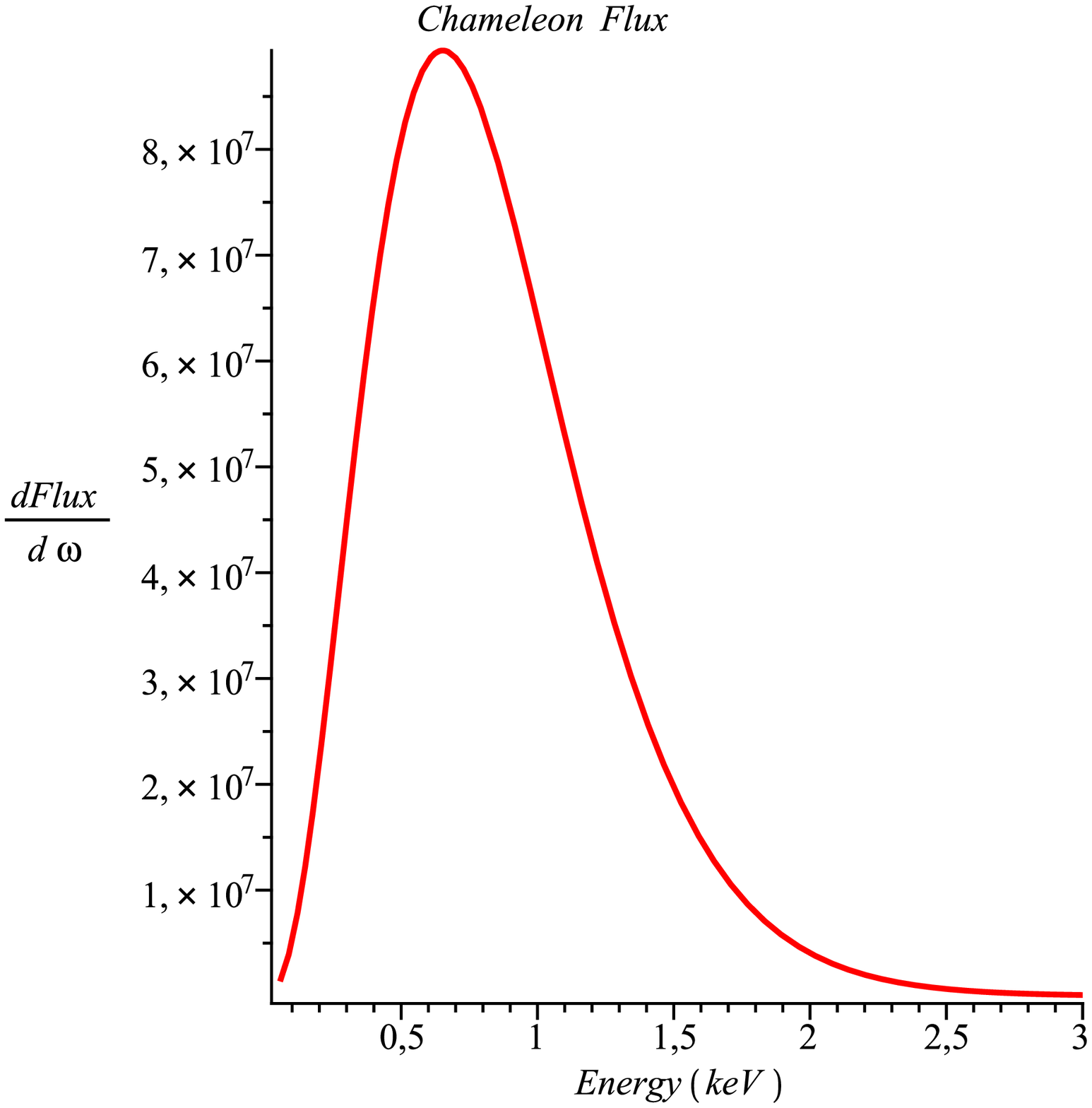}\hspace*{1cm}
\includegraphics[width=7cm]{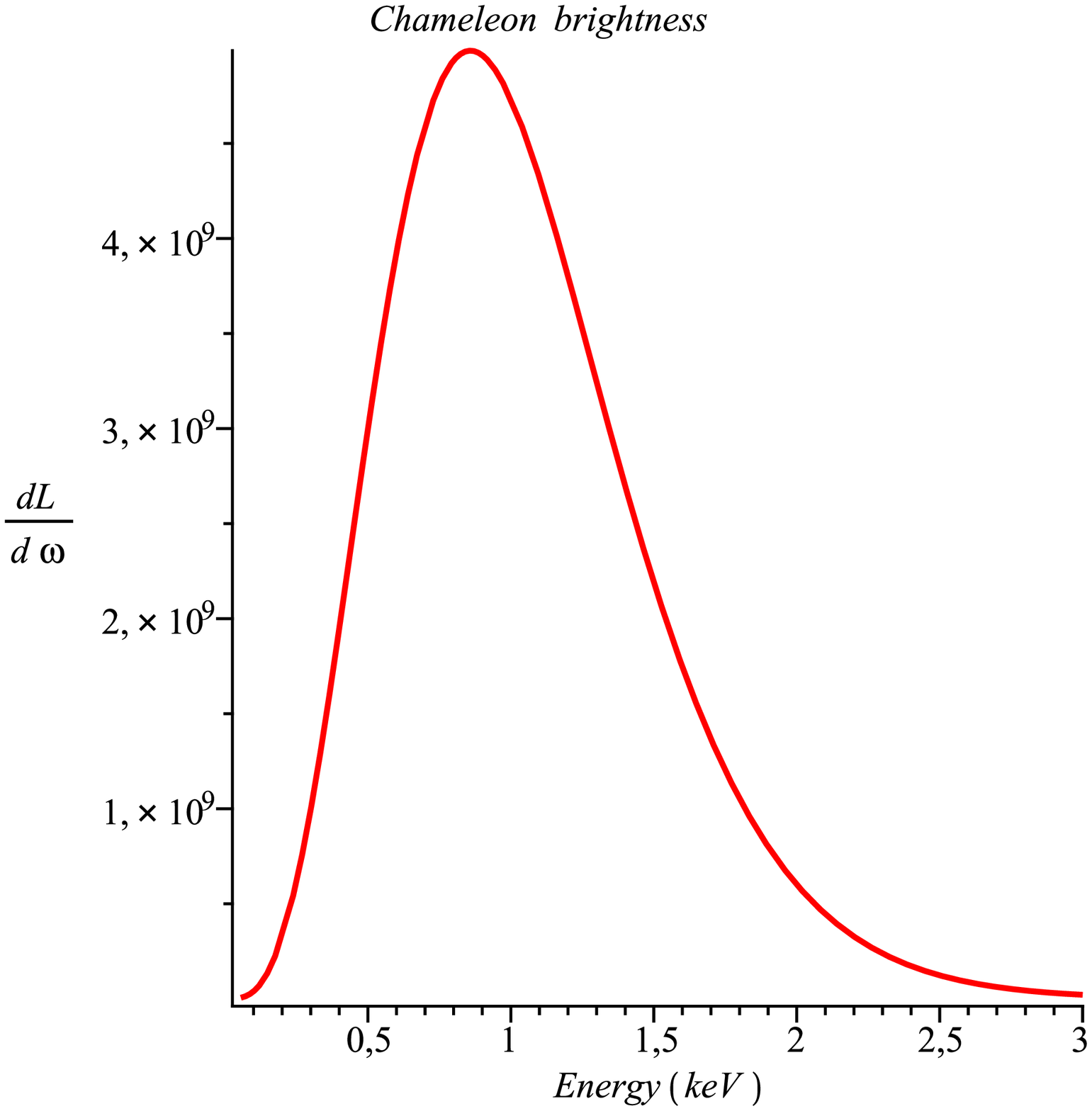}
\caption{On the left, the solar chameleon flux in ${\rm keV^{-1}\cdot s^{-1}\cdot {cm^{-2}}}$ at 1 AU  as a function of the chameleon energy in keV .  Here we have chosen the matter coupling to be $\beta=10^{6}$, the photon coupling $\beta_{\gamma}=10^{10.29}$ and a shell magnetic field in the sun of 30T over a width of 0.01 solar radius above the tachocline. On the right,
the chameleon brightness in ${\rm erg\cdot  keV^{-1}\cdot s^{-1}\cdot {cm^{-2}}}$ at the surface of the sun  as a function of the energy in keV .   The estimated total energy loss in chameleons is $6\cdot 10^9$ ${\rm erg\cdot s^{-1}\cdot {cm^{-2}}}$, i.e. around 10 \% of the photon solar luminosity.}
\end{center}
\end{figure}

The conversion probability  in the helioscope pipe for a small mixing angle $\theta_{\rm helio}$ is
\begin{equation}
P_{\rm helio }(\omega)=  4\theta_{\rm helio}^2 \sin^2 {\Delta}_{\rm helio}
\end{equation}
where $\Delta_{\rm helio}= m_{\rm eff}^2 L/4\omega$ depends on the effective chameleon mass in the helioscope pipe.
When $\Delta_{\rm helio} \ll 1$, the probability is constant and given by $B^2_{\rm helio} L^2/4M^2_{\gamma}$. As $\Delta_{\rm helio}$ increases and as soon as the oscillations due to $\sin \Delta_{\rm helio}$ appear, the probability becomes proportional to $\omega^2 \sin^2 \Delta_{\rm helio}$ which explains why the envelope of the probability distribution is parabolic for very small values of $\omega$ and resembles a sinusoid for larger values before reaching an asymptotic value (not shown).

We are also very interested in the number of chameleons which can be  back-converted to photons in the magnetised photosphere and above. We have calculated the corresponding photon spectra in the photosphere, for the quiet sun outside the photosphere and for flares.
As can be seen in figure 10, the back-converted photon luminosity in the photosphere is negligible. For flaring conditions  at relatively low altitudes, it  is given in figures 10 and 11. In these cases, the coherence length is small and an average over the size of the magnetic regions has been taken.
For flaring conditions at altitudes higher than 10000 km, the back-converted X-ray luminosity can be  rather large. A very interesting result is obtained for the quiet sun since  the back-converted photon spectrum peaks at  $\sim$  1 keV which could explain why the solar corona is even hotter above active regions, see figure 12.

\begin{figure}
\begin{center}
\includegraphics[width=7cm]{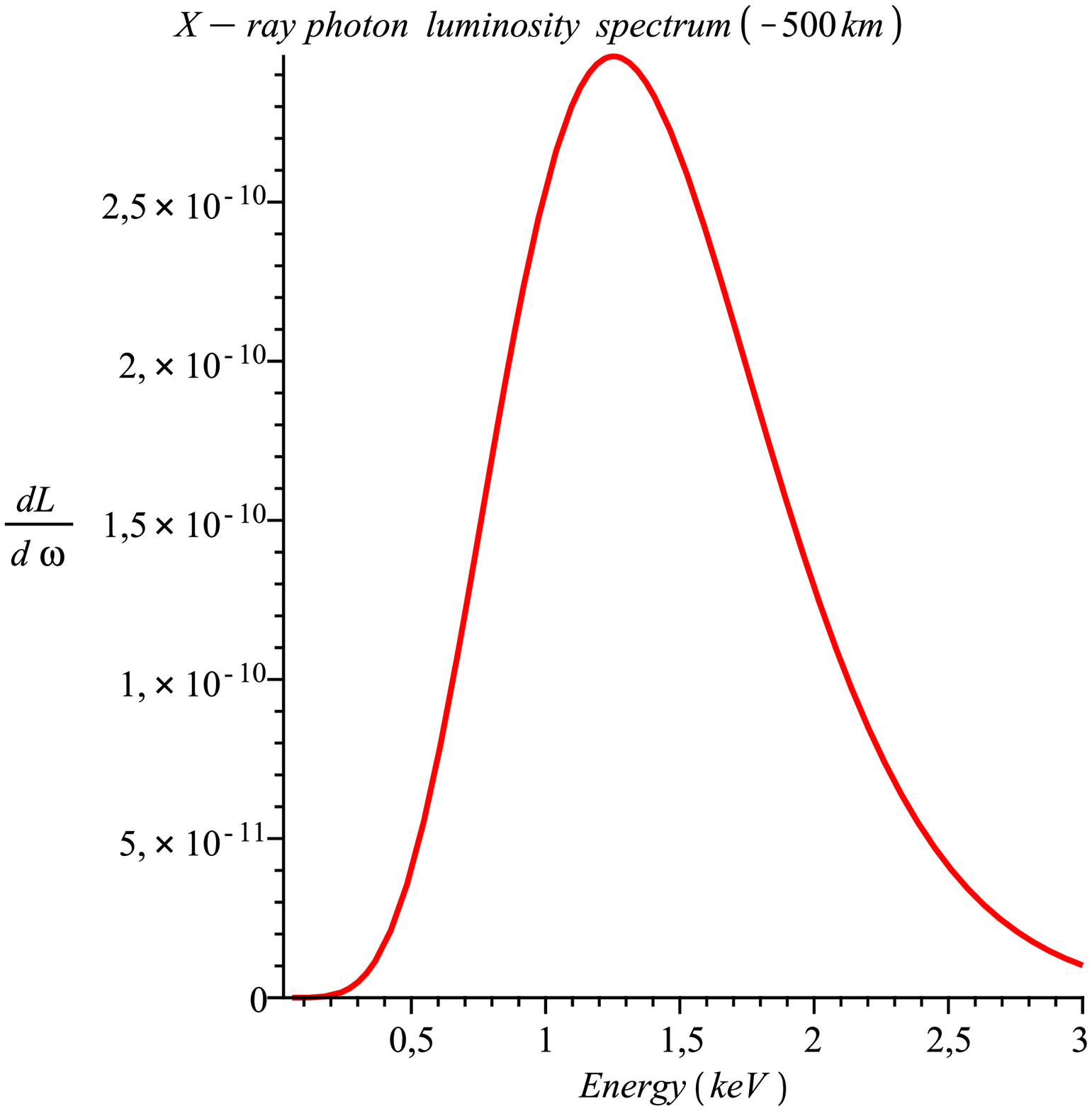} \hspace*{1cm}
\includegraphics[width=7cm]{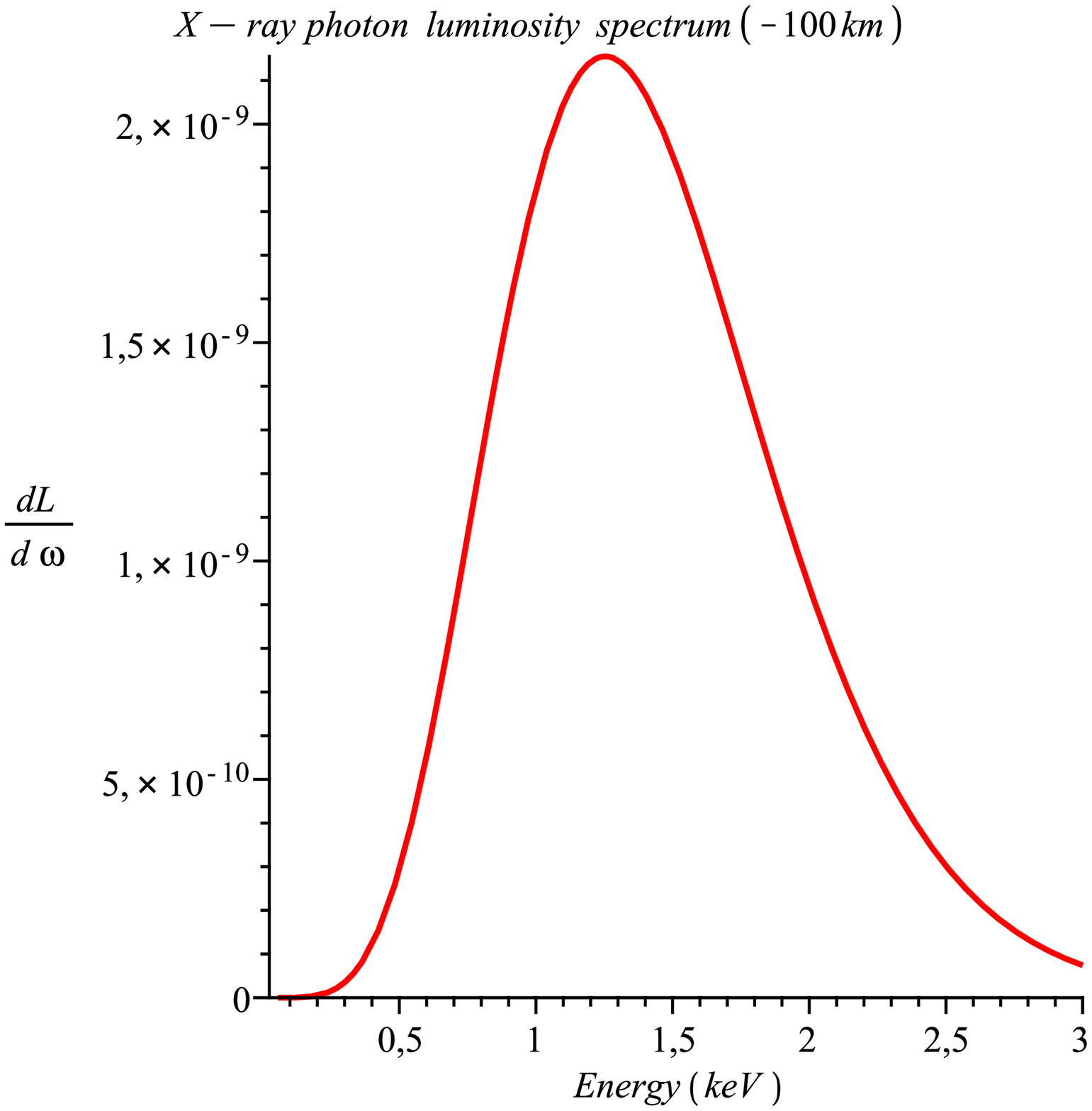} \vspace*{1cm}\\
\includegraphics[width=7cm]{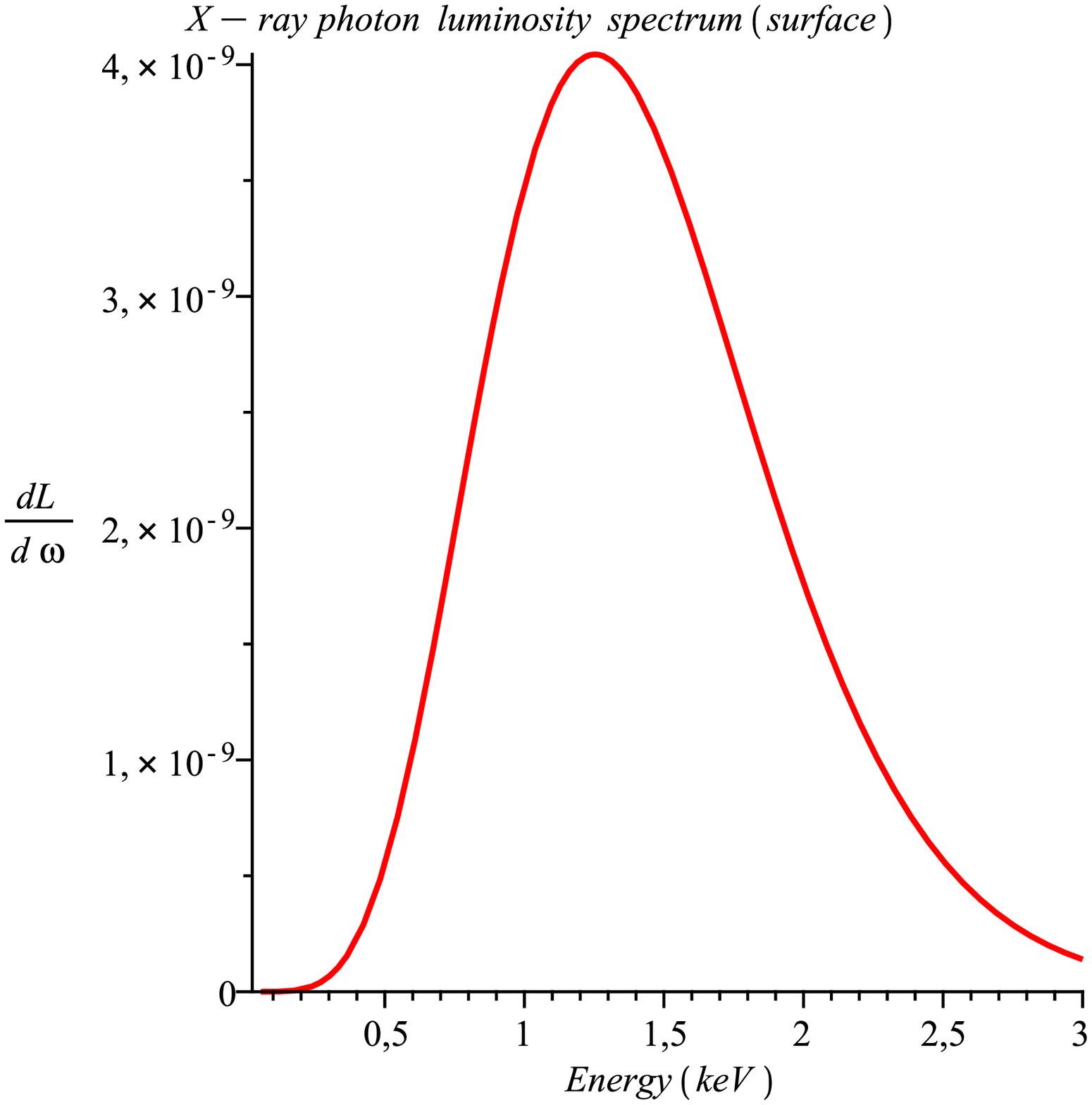} \hspace*{1cm}
\includegraphics[width=7cm]{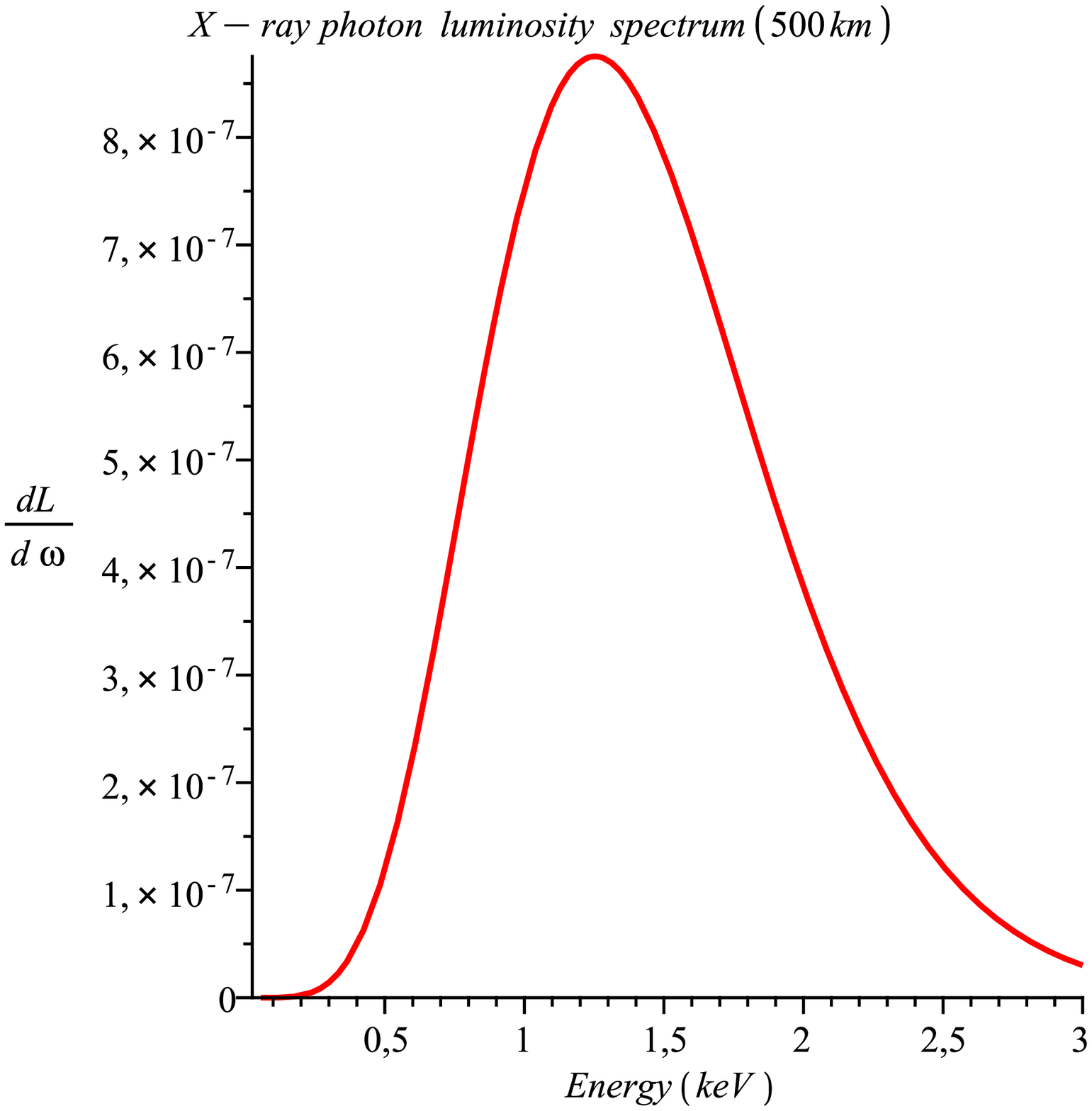}
\caption{The X-ray luminosity spectrum  of back-converted chameleons in keV (in ${\rm erg\cdot keV^{-1}\cdot s^{-1}\cdot {cm^{-2}}}$) leaving the sun as a function of the photon energy . The photons are backconverted at  -500 km, -100 km, 0 km and +500km with magnetic fields 0.2 T in the first three cases and 0.1 T in the last case. In the first three cases, the conversion takes place over one mean free path of order 15 km, 40 km and 50 km while in the last case the conversion takes place over the size of the magnetic region around 100 km. In each case, the coherence length is much smaller than the size of the chameleon to photon  interaction region and the spectrum has been averaged out over many scatterings.
 Here we have chosen the matter and photon coupling to be $\beta=10^{6}$ and  $\beta_\gamma=10^{10.29}$ respectively.}
\end{center}
\end{figure}

\begin{figure}
\begin{center}
\includegraphics[width=7cm]{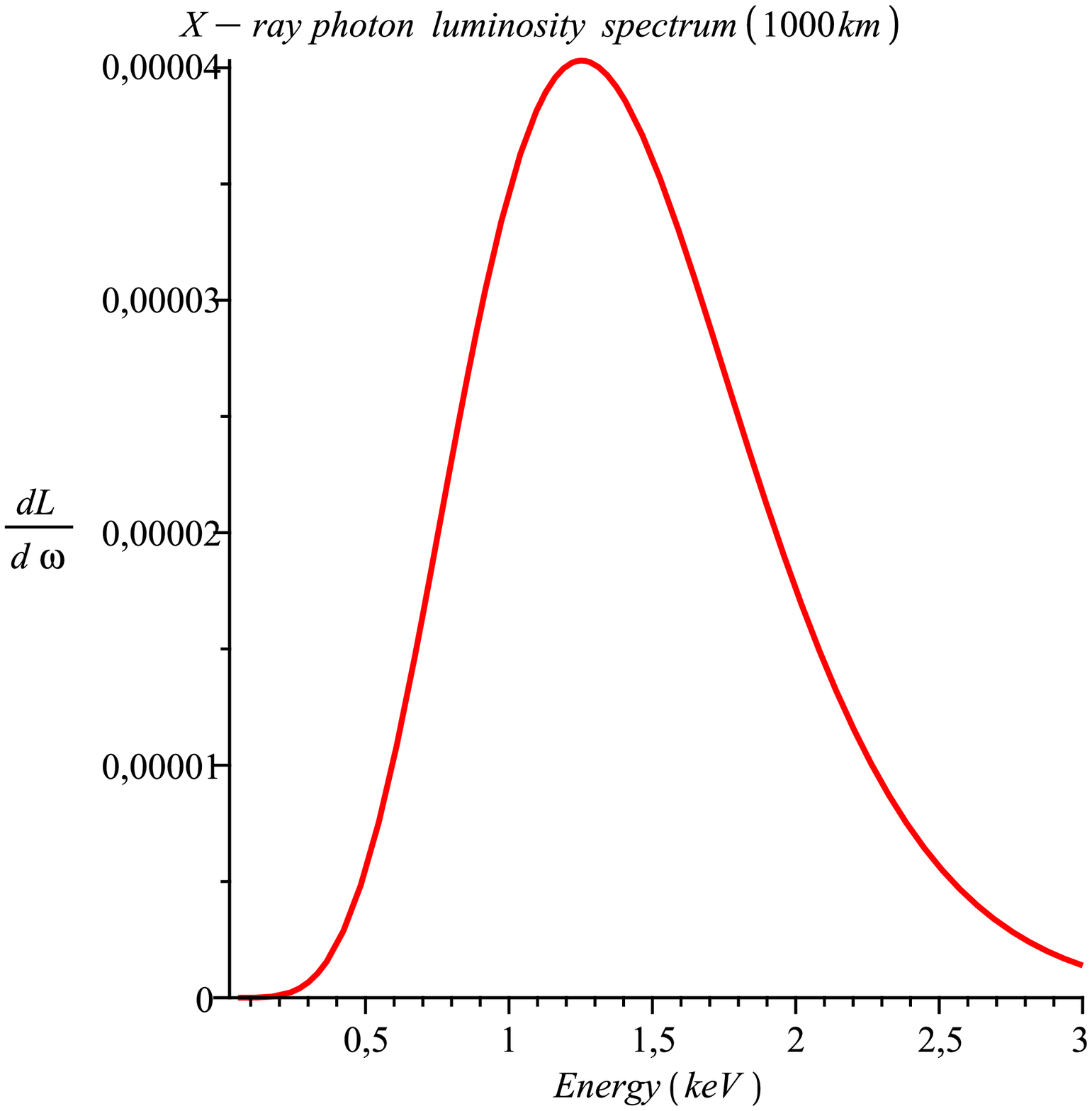} \hspace*{1cm}
\includegraphics[width=7cm]{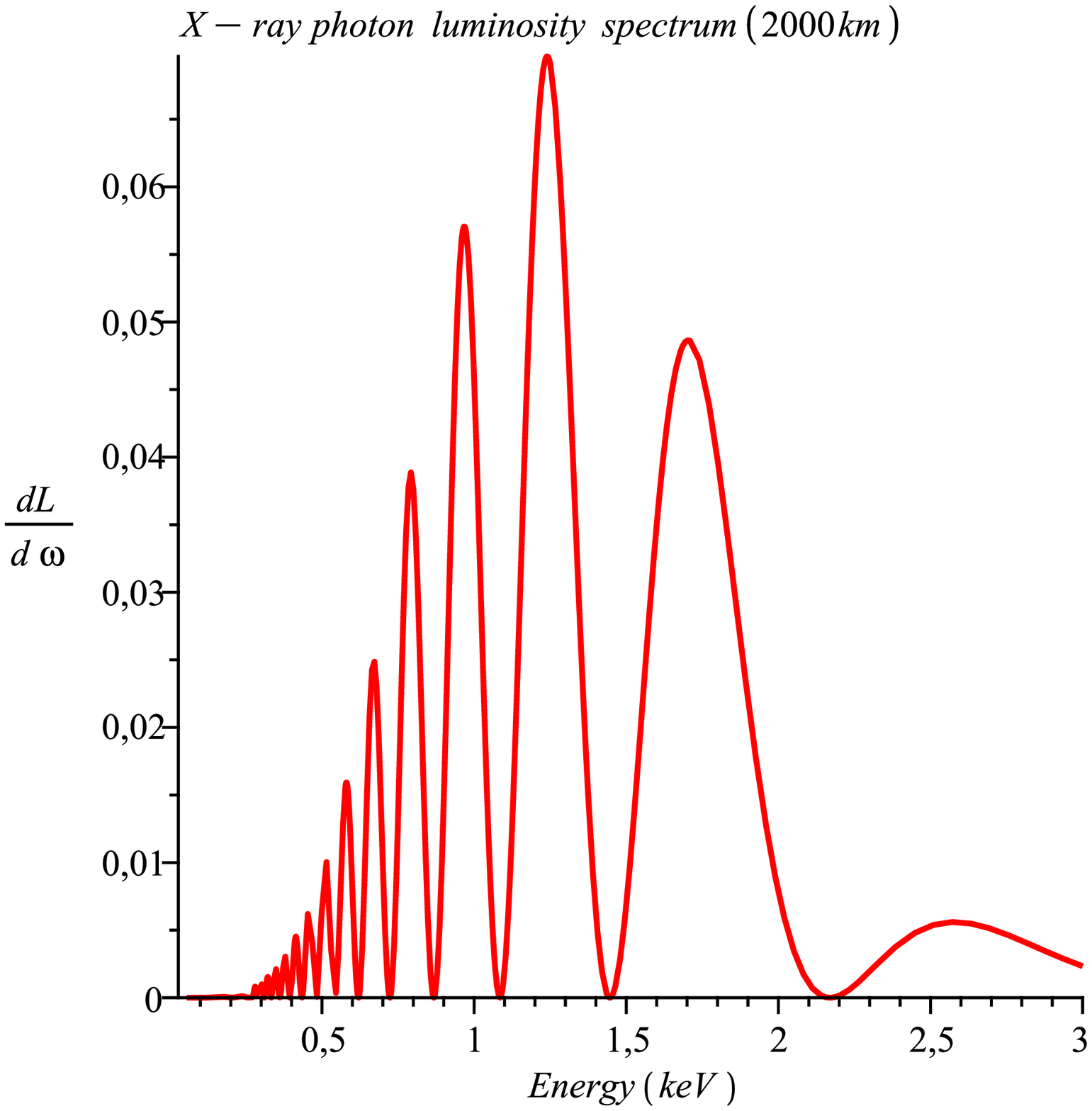} \vspace*{1cm}\\
\includegraphics[width=7cm]{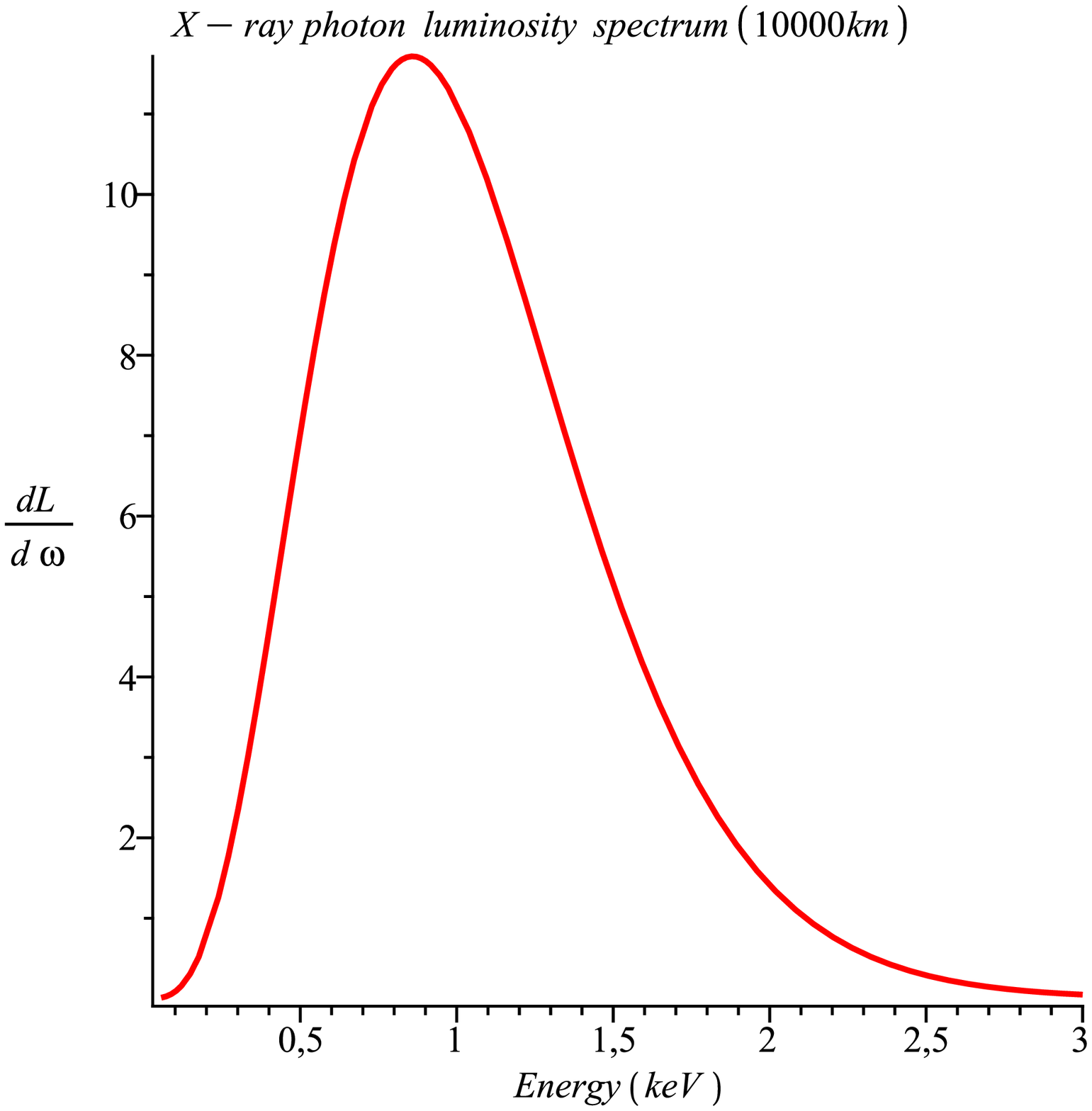} \hspace*{1cm}
\includegraphics[width=7cm]{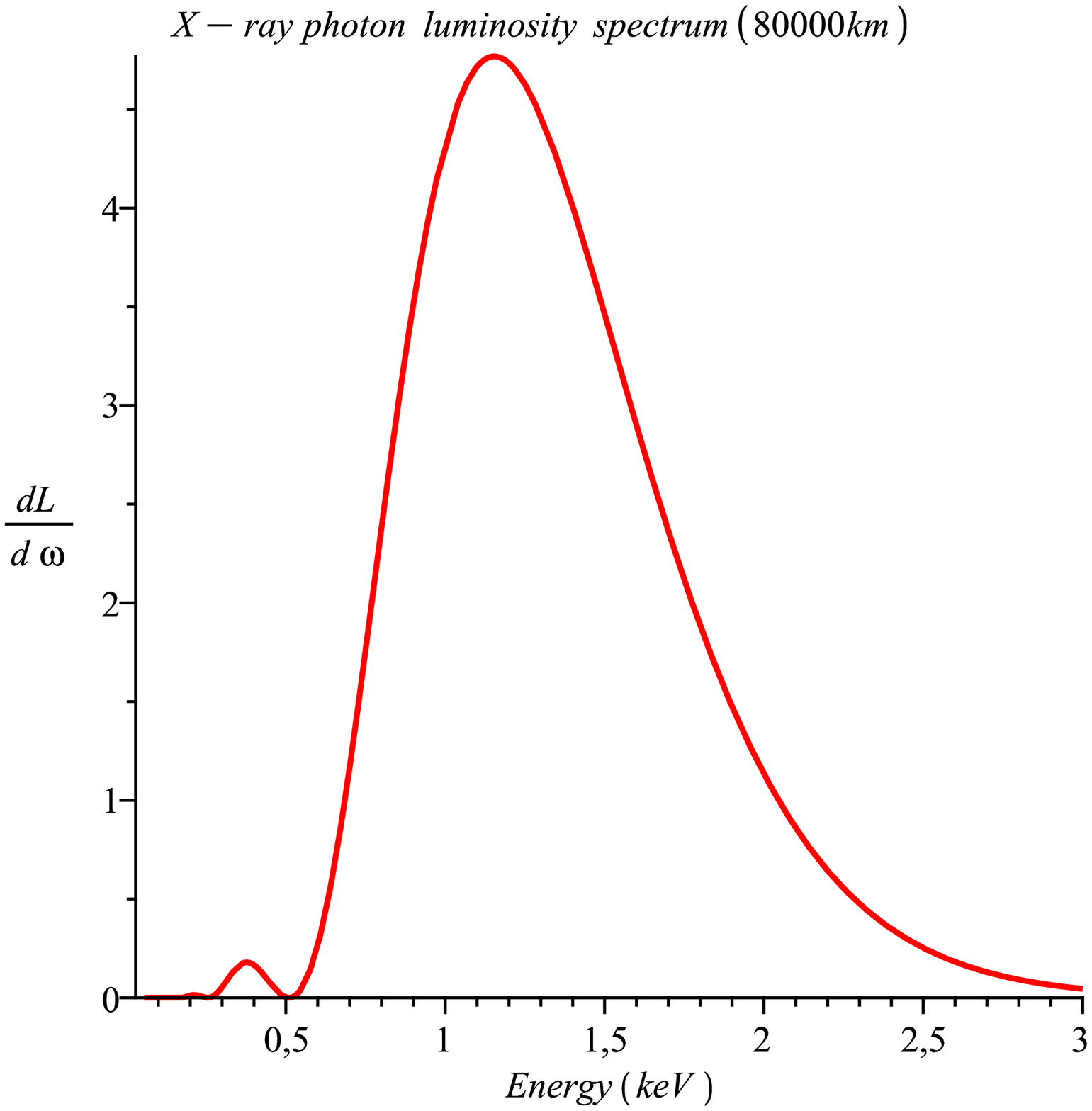}
\caption{The photon luminosity spectrum from  back-converted chameleons (${\rm erg\cdot keV^{-1}\cdot s^{-1}\cdot {cm^{-2}}}$) in solar flares leaving the sun as a function of the photon energy in keV .  The photons are backconverted at  +1000 km, +2000 km, +10000 km and +80000km  with magnetic fields  100 Gauss , 10 Gauss, 10 Gauss  and 5 Gauss respectively. At  the altitude of 1000 km, the spectrum has been averaged out over many oscillations. In the other two cases, we have represented the spectrum itself. In each case we have used  $B_{\rm out} L_{\rm out}\approx 10^4 \ {\rm T.m}$ which determines the length of the magnetic region. Here we have chosen the matter coupling to be $\beta=10^{6}$ and the photon coupling is $\beta_\gamma=10^{10.29}$.
}
\end{center}
\end{figure}

\begin{figure}
\begin{center}
\includegraphics[width=7cm]{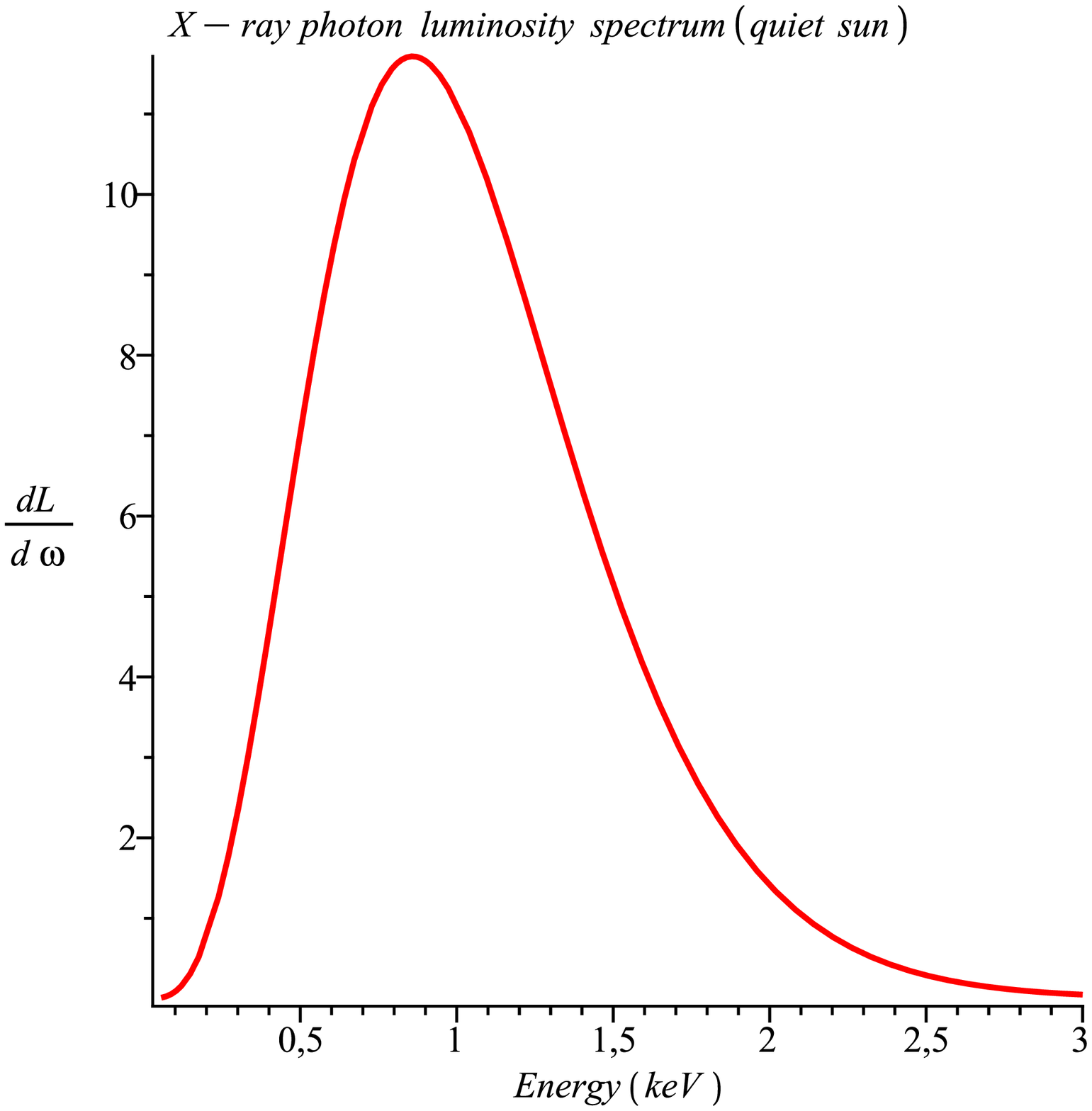}\hspace*{1cm}
\includegraphics[width=7cm]{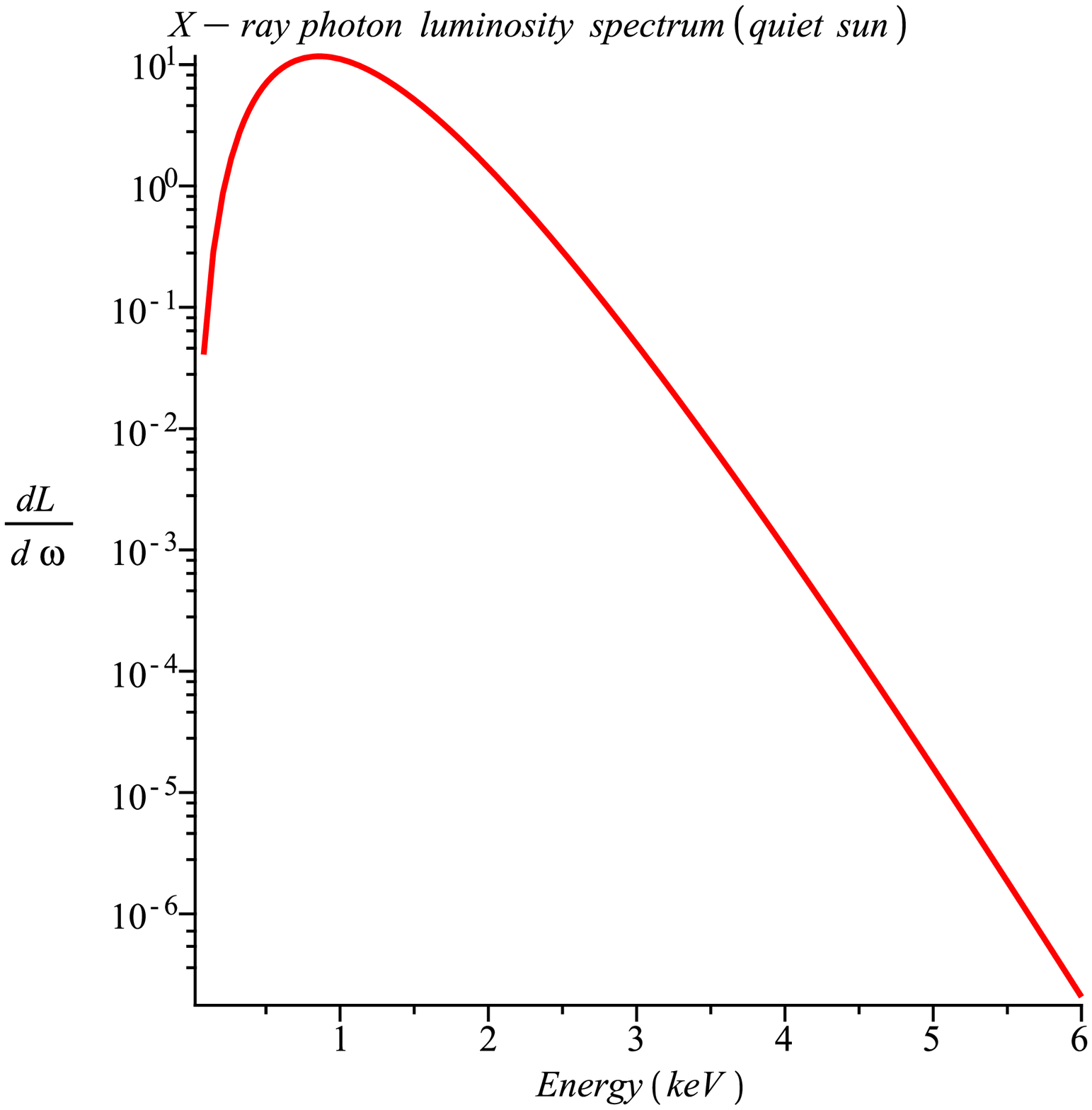}
\caption{The X-ray   luminosity spectrum from  back-converted chameleons leaving the quiet sun as a function of the photon energy in keV in ${\rm erg\cdot keV^{-1}\cdot s^{-1}\cdot {cm^{-2}}}$ (linear scale on the left and logarithmic scale on the right). The photons are backconverted at altitudes larger than 5000 km with a magnetic field  of 1 Gauss and a density of order $10^{-16}{\rm g\cdot cm^{-3}}$.  We have used $B_{\rm out} L_{\rm out}\approx 10^4 \ {\rm T.m}$ to determine the length of the magnetic region. Here we have chosen the matter coupling to be $\beta=10^{6}$ and the photon coupling is $\beta_\gamma=10^{10.29}$.
}
\end{center}
\end{figure}

\subsection{Chameleons through a wall}

The spectrum intensity of regenerated photons depends crucially on the source flux in the X-ray band. We have chosen as a template the  DESY spectrum\cite{desy2} with a total number of $10^{18}\ {\rm s^{-1}\cdot cm^{-2}}$. An upgraded version with a spectrum of $10^{19}\ {\rm s^{-1}\cdot cm^{-2}}$ could be assumed for a future dedicated beam line.
We will quote results obtained with this  flux. The detection of a low flux of  regenerated photon depends also on the detector sensitivity. We will assume in this work $N_{\rm detector}=10^{-6} {\rm s^{-1}}$. With these specifications, we estimate the 2$\sigma$ detection efficiency. For that, we refer to the case $\beta=10^6$ and  $\beta_\gamma=10^{10.32}$ as fixed by the solar energy loss bound.  The original  source spectrum and  the regenerated photons  are presented in figure 13 with $N_{\rm signal}= 6\cdot 10^{-7}{\rm s^{-1}}$.
The number of seconds for a n$\sigma$ detection is
\begin{equation}
S= n^2 \frac{(2N_{\rm detector} + N_{\rm signal})}{N_{\rm signal}^2}
\end{equation}
We find that
a 2$\sigma$ level signal would take 300 days. An increase of the number of photons originating from the source by one or two orders of magnitude would reduce this number to a few days only.
However, if three equivalent CAST magnetic pipes (e.g two LHC bending magnets on each side of the barrier) were used on both sides of the thick barrier, a 5$\sigma$ level would take only 6 days, which is an experimental challenge for a chameleon smoking gun.
A distinguishing feature of chameleons is the fact that there would be no signal on top of the detection noise when the pipe is filled with a gas at relatively high pressure. A comparison between the vacuum and pressured cases inside the magnetic pipes up to a few mbar only may help to identify the regenerated photon signal in a much better fashion.

\section{Conclusion}
We have studied the creation of chameleons inside the sun by the Primakoff effect and their back-conversion to soft X-ray photons in the magnetised photosphere, the corona and  helioscopes on earth by the inverse Primakoff effect.  The production of chameleons is strongly dependent on the assumed features of the magnetised regions inside the sun. Depending on the magnetic field in the sun and the size of the magnetised zone, it is possible to deduce the energy loss of the sun escaping as chameleons and impose the energy loss bound: no more than 10 \% of the
total solar luminosity can be carried away by unknown species. Provided this bound is satisfied, escaping chameleons reach the surface of the earth and may be back-converted into soft X-ray photons in the pipe of a magnetic helioscope. Moreover, a tiny fraction of the escaping chameleons could also be converted into photons inside the photosphere or the corona.

We have found that for a rather conservative  solar model with a widely accepted magnetised region of size $0.01 R_\odot$ around the tachocline, the non-resonant production of chameleons would lead to a 2$\sigma$ detection of X-rays in a CAST-like helioscope and an energy loss of the sun due to chameleons just below 10\%. However, using detectors sensitive to sub-keV photons, this  would   enormousy improve the detection significance.

For the same chameleon parameters, we have proposed a shining through wall experiment where chameleons produced in one or more  CAST-like pipes from a powerful X-ray source
would go through a thick barrier and regenerate into X-rays downstream of the barrier in a second magnetic field. The observation of this phenomenon would be a chameleon smoking gun with clear possible chameleon identification.

In summary we have shown that a viable candidate for dark energy, the chameleon, could be sought for in magnetic helioscopes and purely laboratory based experiments.
Both approaches could be pursued in the near future. The combination of their results could give new insights into yet unexplained phenomena. We have also seen that chameleons could shed new light on the solar corona problem. Of course, nothing guarantees that the coupling of chameleons to photons is large enough to lead to a detection of chameleons. If no signal were obtained, new bounds on this coupling would ensue  corresponding to higher and higher limits on the suppression scale of the coupling of photons to a scalar field.

\acknowledgements
We would like to thank E. Tzelati for discussions on the resonant case.

\begin{figure}
\begin{center}
\includegraphics[width=9cm]{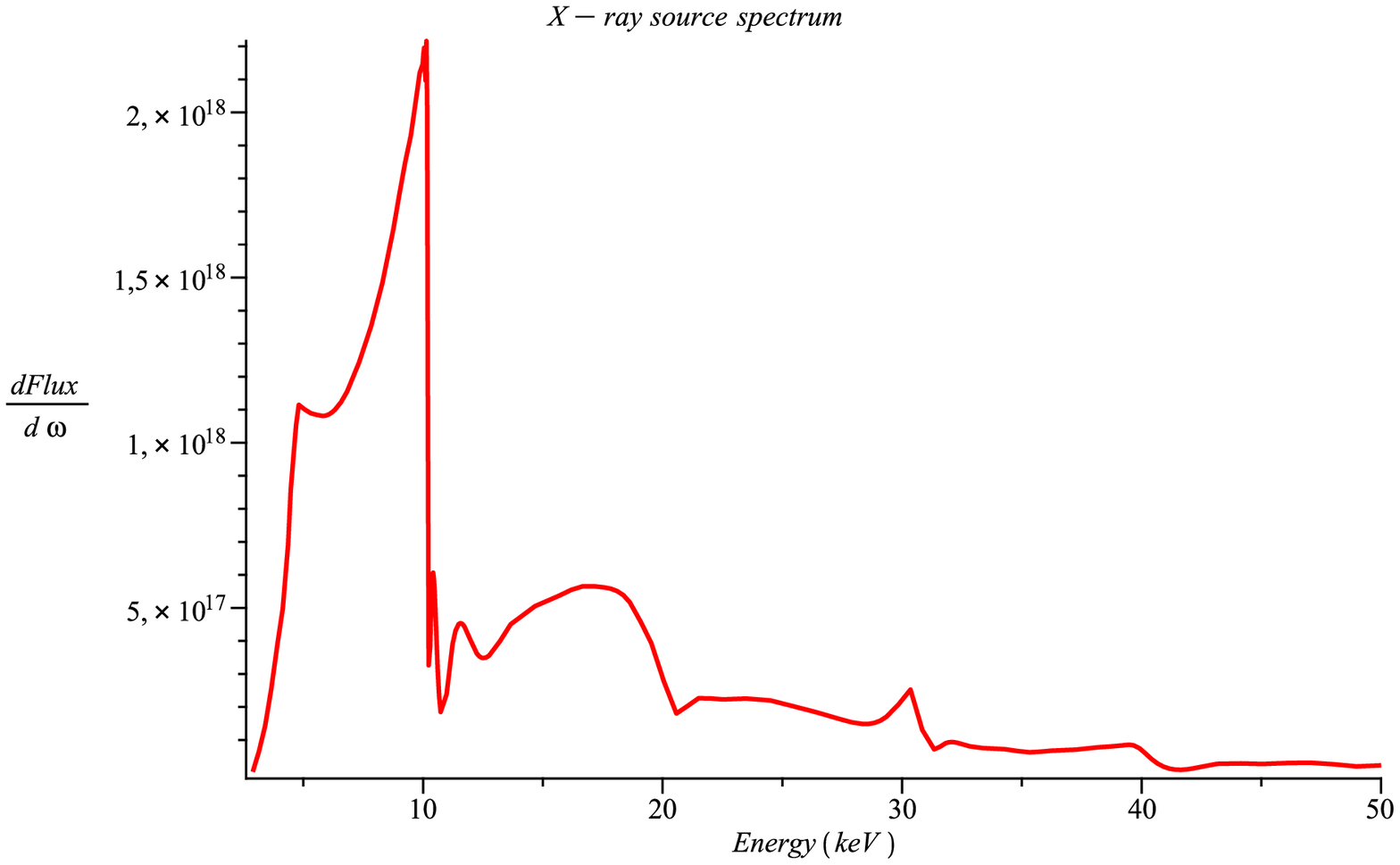}\hspace*{1cm}
\includegraphics[width=9cm]{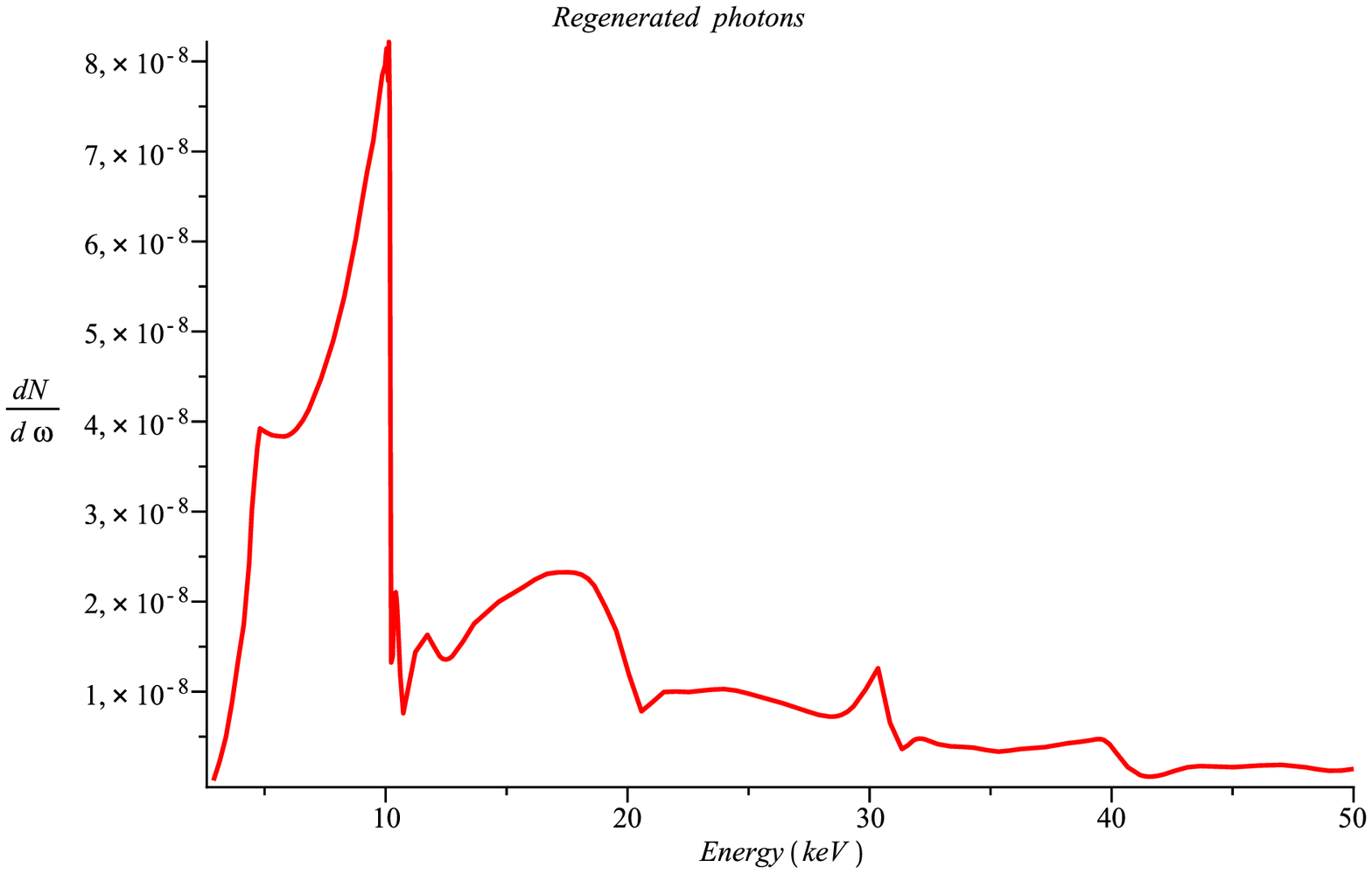}
\caption{The photon flux of the X-ray source for the chameleon-through-a-barrier experiment  in ${\rm cm^{-2}}\cdot{\rm  s^{-1}}\cdot {\rm keV^{-1}}$  as a function of the photon energy in keV (left) and the regenerated spectrum on the other side of the barrier (right). The integrated flux is $10^{19} {\rm s^{-1}\cdot cm^{-2}}$ for the source spectrum. The spectrum of regenerated photons presented here is the one  predicted to be seen by  a chameleon through a wall experiment using two magnetic  pipes on both sides of a thick barrier. The coupling to matter is $\beta=10^6$ while the coupling to photons $\beta_{\gamma}=10^{10.32}$ implies a corresponding  2$\sigma$ result obtained in 300 days with a detection sensitivity of $10^{-6} {\rm s^{-1}\cdot cm^{-2}}$. For magnetic pipes of 3 times the CAST pipe length, a 5$\sigma$ result would only take 6 days.}

\end{center}
\end{figure}

\end{document}